\documentclass[10pt,aps,a4paper,prd,floatfix,onecolumn,notitlepage,nofootinbib]{revtex4-1}

\usepackage[left=2cm,right=2cm,top=2cm,bottom=2cm]{geometry}

\usepackage[utf8]{inputenc}
\usepackage[british,UKenglish,USenglish,american]{babel}
\usepackage{amsmath}

\usepackage{graphicx}
\usepackage{subcaption}

\usepackage{slashed}

\usepackage{mathrsfs}
\usepackage{amssymb}
\usepackage{bbm}

\usepackage{nicefrac}

\usepackage{microtype}

\usepackage{physics}

\usepackage{xcolor}

\usepackage{pifont}

\newcommand{\DD}[1]{ \mathcal{D}#1 \,}

\newcommand\numberthis{\addtocounter{equation}{1}\tag{\theequation}}

\DeclareMathOperator{\arctg}{arctg}

\renewcommand{\bar}{\overline}

\usepackage{bm}\renewcommand{\vec}{\bm}

\def\P{\texttt{+}}
\def\M{\texttt{-}}

\newcommand{\fermi}{\mathrm{F}}

\newcommand{\CPV}{\mathrm{p.v.}}

\newcommand{\sumeta}{\sum_{\eta=\pm 1}}
\newcommand{\sumMatsubara}{\sum_{n=-\infty}^{\infty}}

\newcommand{\cmark}{\ding{51}}%
\newcommand{\xmark}{\ding{55}}%

\begin{document}

\title{
One-meson-loop NJL model: Effect of collective and noncollective excitations on the quark condensate at finite temperature
}

\author{Renan Câmara Pereira}
\email{renan.pereira@student.uc.pt}
\affiliation{Centro de Física da Universidade de Coimbra (CFisUC), 
Department of Physics, University of Coimbra, P-3004 - 516  Coimbra, Portugal}

\author{Pedro Costa}
\email{pcosta@uc.pt}
\affiliation{Centro de Física da Universidade de Coimbra (CFisUC), 
Department of Physics, University of Coimbra, P-3004 - 516  Coimbra, Portugal}

\date{\today}

\begin{abstract}

We explore the effect of including quantum fluctuations in the two flavor Nambu$-$Jona-Lasinio model at finite temperature. This is accomplished, in a symmetry preserving way, by including collective and noncollective modes in the one-meson-loop gap equation which originate from poles and branch cuts in the complex plane, respectively. The inclusion of a boson cutoff, $\Lambda_b$, is necessary to regularize the meson-loop momenta. This new parameter is used to study the influence of going beyond the usual mean field approximation in the quark condensate. As the temperature increases, chiral symmetry tends to get restored, the collective modes melt and only noncollective modes contribute to the quark condensate. With the inclusion of such modes, the
quark condensate at finite temperature has a different behavior than at mean field level that will be explored.
\end{abstract}

\maketitle

\section{Introduction}
\label{introduction}

Introduced by Yoichiro Nambu and Giovanni Jona-Lasinio in 1961, before the assertion of quantum chromodynamics (QCD) as the theory of strong interactions, the Nambu$-$Jona-Lasinio (NJL) model had its debut as a model of nucleons \cite{PhysRev.122.345,PhysRev.124.246}. In the original model, the nucleon fields interact locally to generate the mass gap in the Dirac spectrum, in analogy with the Bardeen-Cooper-Schrieffer theory of superconductivity.
After the establishment of QCD as the theory of strong interactions, the nucleon field was substituted by a quark field \cite{Eguchi:1976iz,Kikkawa:1976fe}. Since then, this model has been widely used as an effective model of QCD, as a result of sharing all the global symmetries of strong interaction, while providing a mechanism for spontaneous breaking and restoration of chiral symmetry. Several improvements have been done to the model throughout the years like the inclusion of finite quark current quark masses \cite{Volkov:1982zx,Ebert:1982pk,Volkov:1984kq}, extending the model for several quark flavors and adding six-quark and eight-quark interactions to better reproduce the hadron spectra \cite{Osipov:2005tq,Osipov:2006ns,Osipov:2007mk}. One of the most important extensions of the model was the inclusion of the Polyakov loop by K. Fukushima \cite{Fukushima:2003fw}. This improvement allowed the incorporation in the model of the ability to describe statistical deconfinement with the spontaneous breaking of $Z\qty(N_c)$ symmetry at finite temperature \cite{Ratti:2005jh,Hansen:2006ee}.

Some examples of the application of this model and its improved versions include hadron phenomenology and behavior at finite temperature and/or density, modeling neutron star matter and study of the QCD phase diagram \cite{Rehberg:1995kh,Costa:2019bua,Costa:2004db,Pereira:2016dfg}. Understanding the phase diagram of QCD is one of the most challenging and interesting topics in modern physics. The experimental study of the QCD phase diagram is one of the major goals of ongoing heavy ion collision (HIC) experiments.  Current experiments like J-PARC in Japan, RHIC at the Brookhaven National Laboratory and SPS at CERN, are not only trying to map the chiral and deconfinement phase boundaries of QCD but also to study the properties of the quark-gluon plasma \cite{Brambilla:2014jmp}. Another goal of these experiments is to characterize the nature of the chiral symmetry restoration and look for the possible existence of the critical endpoint (CEP), predicted by several model calculations. The available tools to theoretically study the QCD phase diagram are limited due to the nonperturbative nature of the theory at low energies. The NJL model has been an important tool to study the phase diagram under different scenarios, for more information on the NJL model and its applications see the reviews \cite{Klevansky:1992qe,Vogl:1991qt,Buballa:2003qv,Hatsuda:1994pi}.

In most of these applications, the NJL was studied in the standard mean field (MF) approximation, equivalent to the so-called Hartree plus random phase approximation on the quark polarization function (RPA) \cite{ZHUANG1994525}. Within these schemes, only the quark loop is considered at the effective action level and quantum fluctuations caused by meson modes are neglected. Including fluctuations in the NJL model, however, is not an easy task \cite{ZHUANG1994525}. For some works, including beyond MF corrections to the NJL model and linear sigma model see \cite{Guo_1997,Dubinin:2016wvt,Fuseau:2019zld,Blaschke:2013zaa,Jiang:2010rn,RONER2008118,Hell:2009by,Blaschke:2014zsa,Nemoto:1999qf,Baacke:2002pi,Petropoulos:2004bt,Baacke2004,Andersen:2008qk}.

Studying the model beyond the MF approximation, is very important to correctly inspect the physical behavior near the critical region where the system displays long range correlations. At low temperatures and densities, before the restoration of chiral symmetry, it is expected that a major role is played by the thermal excitations of the pion modes \citep{Florkowski:1996wf}. These low mass degrees of freedom are the pseudo-Goldstone modes of the NJL model. Including quantum fluctuations in QCD effective model calculations, is also known to smooth the transition and bring the critical region toward lower values of temperature \cite{Haas:2013qwp,Morita:2011jva}. The localization of the critical region in model calculations can be essential to aid experimental efforts to pinpoint the CEP.

One widely used technique to include quantum fluctuations beyond the MF, is the functional renormalization group (FRG). The FRG is a powerful nonperturbative method that incorporates the Wilsonian idea of a gradual momentum integration. In this method, the central object is the, scale-dependent, average effective action which acts as an interpolation functional between two regimes: the ultraviolet scale, without quantum fluctuations and the infrared scale, where all quantum fluctuations have been taken into account. For reviews on the FRG see \cite{Gies:2006wv,Pawlowski:2005xe}. The FRG has been applied to the NJL model with a scale dependence incorporated in the four-Fermi interaction coupling. However, such a scheme leads to a diverging coupling in the renormalization group flow, that signals spontaneous breaking of chiral symmetry, for more information see \cite{Braun:2011pp,Aoki:2014ola,Fukushima:2012xw}. Recently, different schemes have been used to apply the FRG to a theory with four-Fermi interactions, like the NJL model, see \cite{Braun:2011pp,Aoki:2014ola,Fukushima:2012xw,Aoki:2015hsa,Aoki:2015mqa}. The FRG technique was also successfully applied to the quark-meson model to study the QCD phase diagram in both two and three flavors, as well as used to compute spectral functions through a simple analytical continuation to imaginary time \cite{Schaefer:2004en,Tripolt:2013jra}. However, more recently, it was also found that the application of the FRG to the 2-flavor QM model leads to an unphysical behavior at low temperatures and high chemical potentials: the existence of a region of negative entropy density near the first-order phase transition of the model, for more information see \cite{Tripolt:2017zgc}.

Any calculation scheme used to solve the model, at any level of approximation, must be symmetry conserving, i.e., it has to preserve the symmetries of the model. In the case of chiral symmetry and its breaking in the vacuum, the model must have a Goldstone mode in the chiral limit. The MF or Hartree plus RPA calculations are symmetry preserving \cite{ZHUANG1994525,HUFNER1994225}.

Different symmetry conserving schemes, to take the NJL model beyond the MF approximation, have been presented over the years like the $1/N_c$ expansions, ``$\Phi-$derivable'' methods  and functional methods \cite{Nikolov:1996jj,Muller:2010am,ZHUANG1994525,Oertel:2000jp,Plant:2000ty,Blaschke:1995gr,YAMAZAKI201319}. The MF approximation represents the leading order in the $N_c$ expansion and corrections to the MF could be of order $1/N_c$. However, the NJL is a nonrenormalizable field theory in four spacetime dimensions. A regularization procedure must be applied, which will be directly related to the absolute size of the corrections. This means that the magnitude of the corrections is not only dictated by the expansion parameter \cite{Buballa:2003qv} but also by the model parametrization and phenomenology.

A nonperturbative and symmetry conserving method was developed by E. Nikolov \textit{et al.} in \cite{Nikolov:1996jj}, based on the effective action formalism. Such formalism was coined as the one-meson-loop approximation \cite{Nikolov:1996jj,Oertel:2000jp} and represents the next to leading order correction in the $N_c$ expansion of the NJL effective action. It was later extended to include low temperature effects in the gap equation \cite{Florkowski:1996wf} using an approximation: at low temperatures, only the lowest lying pion pole would contribute.

In this work, we do not deal with the system at finite density. Hence, vector-type interactions will not be considered even though these interactions can be present in the NJL model, specially at finite density. Indeed, vector interactions are present if one considers the NJL model as an effective model of QCD, based on the color-current expansion and the Fierz transformation of the one-gluon exchange interaction  \cite{alkofer2008chiral,Garibli2019}.

In the present paper, we will take an important step in the direction of calculating the NJL phase diagram beyond the MF approximation by extending the symmetry conserving scheme presented by E. Nikolov \textit{et al.} to finite temperature \cite{Nikolov:1996jj,Florkowski:1996wf}. To accomplish this, we will solve the NJL gap equation including all contributions coming from the one-meson-loop correction terms. This will allow us to study the impact of meson fluctuations on the quark condensate and in the restoration of chiral symmetry at finite temperature. Previous works have only considered the effect of the one-meson-loop terms in the vacuum quark condensate and did not develop the formalism of the integration technique and phenomenology to extend the calculation to finite temperature \cite{Nikolov:1996jj,Florkowski:1996wf,Oertel:2000jp,HUFNER1994225}. In order to consider all contributions coming from the one-meson-loop correction terms, we will separate the contour integrations that arise in the calculation in two distinct contributions: the collective and noncollective modes \cite{YAMAZAKI201319,YAMAZAKI2014237}. This calculation does not involve meson fields with kinetic boson terms at the Lagrangian level. In this formalism, mesons are composite collective and noncollective excitations of the underlying quark fields.

This paper is organized as follows. In Sec. \ref{model_and_formalism} the NJL model and formalism, to derive the one-meson-loop gap equation, are presented. The separation of the collective and noncollective modes is laid out. In Sec. \ref{results} the results from increasing the meson fluctuations in the vacuum and at finite temperature are studied. The separated effect of the collective modes and noncollective modes, on the quark condensate, is also considered. Finally, in Sec. \ref{conclusions} conclusions are discussed and further work is planned.

We work with units in which $\hbar = c = 1$. The following notation is used along the work for an $n-$dimensional integration in momentum space:
\begin{align*}
\int_{ q_n } & = \int \frac{ \dd[n]{q} }{ \qty(2 \pi)^n } .
\end{align*}

\section{Model and Formalism}
\label{model_and_formalism}

To derive the NJL gap equation including one-meson-loop corrections, we will use the effective action formalism. Following \cite{Nikolov:1996jj}, we consider the two flavor NJL model, whose Lagrangian density in Minkowski spacetime is given by: 
\begin{align}
\mathcal{L} \qty( \bar{\psi}, \psi ) = 
\bar{\psi} \qty( i \slashed{\partial} - \hat{m} ) \psi + 
\frac{g_s}{2}  \qty( \bar{\psi} \hat{\Gamma}_\alpha \psi )^2  .
\label{lagrangian_NJL_SU2}
\end{align}
Here, $\psi$ is the quark field, $\hat{m}$ is the current quark mass, which explicitly breaks chiral symmetry and $g_s$ is the coupling constant of the scalar and pseudoscalar four-fermions interaction. The mass dimension of the coupling  $g_s$ is $\qty[ D-6 ]$ (where $D$ is the spacetime dimension), rendering this model nonrenormalizable in $D=4$. The operator $\hat{\Gamma}_\alpha$, to incorporate both the scalar and pseudoscalar interactions, is defined as $\hat{\Gamma}_0 = \mathbbm{1}$ and $\hat{\Gamma}_j = i \gamma^5\tau^j$.

To include temperature we will use the Euclidean spacetime by performing a Wick rotation from real times to imaginary times, $x_0 \to -i\tau$ changing the metric, $\eta_{a b} = - \delta_{a b}$. The Euclidean action can be written as $\mathcal{S}_E \qty[ \bar{\psi}, \psi ] = - \int_0^{\nicefrac{1}{T}} \dd{\tau} \int \dd[3]{x} \mathcal{L}_E \qty( \bar{\psi}, \psi )$. The generating functional of the fully connected Green’s functions, for a given temperature $\qty(T)$, ignoring a normalization factor, can be written as:
\begin{align}
\mathcal{W} \qty[ T; \eta, \bar{\eta} ] = 
\ln
\int \DD{\bar{\psi}} \DD{\psi}  
e^{ 
-\mathcal{S}_E \qty[ \bar{\psi}, \psi ] + 
\int_0^{\nicefrac{1}{T}} \dd{\tau} \int \dd[3]{x}
\qty( \bar{\psi}\eta + \bar{\eta} \psi  ) 
}  .
\end{align}
When dealing with multiquark interactions one can use the Hubbard-Stratonovich transformation to absorb these nonquadratic interactions with the use of auxiliary fields with the same quantum numbers as the quark bilinears operators. In the case of the NJL model, it can be written as \cite{Ebert:1997fc}:
\begin{align}
\exp \qty{
\int_0^{\nicefrac{1}{T}} \dd{\tau} \int \dd[3]{x}
\frac{g_s}{2} \qty( \bar{\psi} \hat{\Gamma}_\alpha \psi )^2
}
\propto
\int \DD{\phi_\alpha}
\exp \qty{
-\int_0^{\nicefrac{1}{T}} \dd{\tau} \int \dd[3]{x}
\qty[
\frac{\phi_\alpha^2}{2 g_s} + \qty( \bar{\psi} \hat{\Gamma}_\alpha \psi ) \phi_\alpha
]
} .
\end{align}
This exact transformation leads to a partially bosonized version of the model with Yukawa-type of interactions between the fermions and the auxiliary fields without kinetic terms. The quadratic fermionic term can then be integrated out exactly. In this model the zeroth component of the field $\phi_0$, correspond to a scalar  meson field and the other three $\vec{\phi}$, to a pseudoscalar meson field.

After using this transformation and shifting variables as $\phi_\alpha \to \phi_\alpha - m \delta_{\alpha 0}$, the quark fields can be integrated out to yield the completely bosonic energy functional,
\begin{align}
\mathcal{W} \qty[ T; J ] = 
\ln
\int \DD{\phi_\alpha}
e^{ 
-\mathcal{S}_E \qty[ \phi_\alpha ] + 
\int_0^{\nicefrac{1}{T}} \dd{\tau} \int \dd[3]{x}
J_\alpha \phi_\alpha
}  .
\end{align}

As pointed out in \cite{Nikolov:1996jj,ripka1997quarks}, for two quark flavors the complex part of the fermionic determinant vanishes and we are allowed to write $\tr \ln  D = \nicefrac{1}{2} \tr \ln D^\dagger D$. Hence, the bosonic action $\mathcal{S}_E \qty[ \phi ]$, can be written as:
\begin{align}
\mathcal{S}_E \qty[ \phi ] & = 
-\frac{N_c}{2} \tr \ln D^\dagger D + 
\int_0^{\nicefrac{1}{T}} \dd{\tau} \int \dd[3]{x}  
\qty[
\frac{\phi_\alpha^2}{2 g_s} - \frac{ m \phi_0}{g_s}  + \frac{ m^2 }{2 g_s} 
] .
\label{eq:classical_action}
\end{align}
Where the operator $D^\dagger D$ is given by:
\begin{align}
D^\dagger D & = -\partial_a \partial_a + i \gamma_a \hat{\Gamma}_b \qty( \partial_a \phi_b ) + \phi_a \phi_a.
\label{DD_covariant}
\end{align}
Following the Wick rotation, the partial differential operator is defined as $\partial_a = \qty( -i \partial_0, \vec{\nabla} ) = \qty( \partial_\tau, \partial_{\vec{x}} )$ and the Euclidean Dirac matrices are $\gamma_a = \qty( i\gamma^0, \vec{\gamma} ) = \qty( \gamma_\tau, \vec{\gamma} )$, which respect the anticommutation relation $\qty{ \gamma^a, \gamma^b } = -2\delta^{a b}$.

The effective action of the model can be obtained through a Legendre transformation,
\begin{align}
\Gamma \qty[ T; \varphi ] = \int \dd{\tau} \int \dd[3]{x}  J_a \varphi_a - \mathcal{W} \qty[ T; J ] ,
\end{align}
where $\varphi_a$ is the vacuum expectation value of the fields in the presence of an external source,
\begin{align}
\varphi = \expval{ \phi }_J = 
\fdv{ \mathcal{W} \qty[ J ] }{ J } .
\end{align}
Considering small fluctuations around the background field, one can expand the effective action in terms of the action given by Eq.  (\ref{eq:classical_action}) and its functional derivatives \cite{ripka1997quarks}. Using such expansion, the one-meson-loop effective action is: 
\begin{align}
\Gamma \qty[ T; \varphi ] = 
\mathcal{S}_E \qty[ \varphi ] + 
\frac{1}{2} \tr \ln \fdv[2]{ \mathcal{S}_E \qty[ \varphi ] }{ \varphi } .
\label{one-loop-effective-action}
\end{align}
Keeping only the first term corresponds to the mean field approximation.

The NJL one-meson-loop gap equation can be derived by requiring that, for a given constant field configuration, the effective action in Eq.  (\ref{one-loop-effective-action}) is stationary. To respect the symmetries of the vacuum, only the scalar field can have a nonvanishing expectation value, $\bar{\varphi} = \qty(S,\vec{0})$. One writes:
\begin{align}
\left.
\fdv{\Gamma \qty[ \varphi ]}{\varphi_c}
\right|_{ \bar{\varphi} } 
=
\fdv{ \mathcal{S}_E \qty[ \varphi ] }{ \varphi_c }
+
\frac{1}{2} \tr 
\Delta_{ab}
\frac{ \delta^3 \mathcal{S}_E \qty[ \varphi ] }{ \delta \varphi_a \delta \varphi_b \delta \varphi_c}
= 0.
\label{eq:gap_equation_definition}
\end{align}

The first term in the gap equation is the MF contribution while the remaining terms correspond to the contribution coming from the meson fluctuations.

In the MF approximation, the pole of the quark propagator is given by $\varphi_0=S$, meaning that the constituent MF-quark mass $m_\psi$, is given by $m_\psi=S$. As already pointed out by other authors \cite{Nikolov:1996jj,Oertel:2000jp}, the same does not occur on the one-meson-loop calculation and $S$ is no longer identifiable with the quark mass. However, we will continue to call $S$ the Hartree mass, since it can still be interpreted as a mass scale and it is essential on the definition of the masses of collective and noncollective modes that will contribute to the quark condensate, which can be calculated using \cite{Nikolov:1996jj,Oertel:2000jp}:
\begin{align}
\expval{ \bar{\psi} \psi } = - \frac{ \qty( S - m ) }{ g_s } .
\end{align}
Using the interpretation of $S$ as the MF or Hartree quark mass, will be important to understand the behavior of the meson modes with increasing temperature.

The function $\Delta_{ab}^{-1} (S,q)$, needed to solve Eq.  (\ref{eq:gap_equation_definition}), is the second variation of the bosonic action $\mathcal{S}_E \qty[ \varphi ]$, with respect to the fields at the stationary point. It can be calculated to yield:
\begin{align}
\Delta_{ab}^{-1} (S,q) = 
\delta_{ab}
\qty[ 
2 N_c N_f f_1\qty(S,q) \qty(q^2 + 4S^2 \delta_{0 \sigma} )
-4 N_c N_f f_0\qty(S) + g_s^{-1}
]
.
\label{eq:MF_action_second_derivative}
\end{align}
This is the meson propagator in the MF approximation which also agrees with the RPA meson propagator. The functions $f_0\qty(S)$ and $f_1\qty(S,q)$ are so-called quark-loop functions (they are presented in Appendixes \ref{appendix_f0_loop_function} and \ref{appendix_f1_loop_function}, respectively). 

As suggested first by E. Nikolov \cite{Nikolov:1996jj} and after by M. Oertel \cite{Oertel:2000jp}, in order to have a symmetry conserving calculation, ensuring the pion as the Goldstone mode in the chiral limit, the quantity in Eq.  (\ref{eq:MF_action_second_derivative}), in every expression, has to be substituted by: 
\begin{align}
\tilde{ \Delta }_M^{-1}  (S,q) & = 
2 N_c N_f f_1\qty(S,q) \qty(q^2 + 4S^2 \delta_{M\sigma} ) + \frac{m}{g_s S} ,
\label{eq:tilde_meson_propagator_def}
\end{align}
the meson-loop propagator, yielding the so-called meson-loop-approximation (with $M=\qty{ \sigma, \pi }$). This substitution is exact in the MF approximation, where the MF-gap equation ensures its validity. In the first derivation of the one-meson-loop gap equation by E. Nikolov \textit{et al.} \cite{Nikolov:1996jj}, this substitution is justified in the basis of an $N_c$ counting scheme. The first term in the gap equation (the quark loop term) is of order $N_c$. The second term will be of order $1/{N_c^0}$. Using the definition given in Eq.  (\ref{eq:MF_action_second_derivative}) would lead to contributions in the gap equation of order $1/N_c$, introducing higher order corrections in the calculation and ruining the $N_c$ counting scheme. Substituting by Eq.  (\ref{eq:tilde_meson_propagator_def}) makes the calculation consistent and leads to massless pion in the chiral limit, as shown by the authors. For more details on their argument, see \cite{Nikolov:1996jj}. This substitution was also employed by M. Oertel \textit{et al.} \cite{Oertel:2000jp}, where it is argued to be necessary in order to make the argument of the logarithm in Eq.  (\ref{one-loop-effective-action}) positive definite, yielding a real and positive solution to the one-meson-loop gap equation. (For more details see Ref. \cite{Oertel:2000jp}). In the present work we use this approximation since it is necessary to have a symmetry conserving approximation when adding meson loops in the effective action formalism. The functions $\tilde{ \Delta }_\sigma(S,q)$ and $\tilde{ \Delta }_\pi(S,q)$ do not correspond to the meson propagators with one-meson-loop corrections. To effectively calculate the meson propagators with one-meson-loop corrections, one would have to calculate the second functional derivative of the effective action including the one-meson-loop term, generating third and fourth order functional derivatives of the bosonic action given in Eq.  (\ref{eq:classical_action}). This is beyond the scope of the present work.

As already stated, the NJL model is nonrenormalizable and some regularization scheme is needed in order to mathematically define the model. Here we will apply a 3-momentum regularization in all momentum integrations, effectively truncating the Hilbert space of the fields \cite{Ripka:1997zb}. The quark loop momentum can be regularized with a hard 3D-momentum cutoff, $\Lambda_f$, the fermion cutoff. When including quantum fluctuations in the calculation, due to the nonrenormalizable nature of the model, a new parameter has to be introduced in order to regularize the meson loops, $\Lambda_b$, the boson cutoff. Trivially, when $\Lambda_b=0$, one recovers the MF approximation. Upon studying the effect of quantum fluctuations beyond the mean field, in the NJL model, several authors have studied the ratio $\alpha=\Lambda_b / \Lambda_f$, or even fixed this ratio to an arbitrary value when building NJL models which dealt with meson loop corrections \cite{Nikolov:1996jj,Oertel:2000jp,Florkowski:1996wf,YAMAZAKI201319,PhysRevC.96.045205,Blaschke:2017pvh}. In this study, we independently choose the values of $\Lambda_f$ and $\Lambda_b$ because the mathematical relation between them are not well determined in the NJL model at present.

In this work we fix the ratio $\alpha=\Lambda_b / \Lambda_f$ by fixing the energy scale of the model. The one-meson-loop contribution has a clear connection with the quark loop term: the mesons in this formalism are composite collective and noncollective excitations of the underlying quark fields and are not meson fields with kinetic boson terms at the Lagrangian level. This is clear from the explicit dependence on the $f_1\qty(S,q)$ loop function in the one-meson-loop terms. In fact, the largest energy in the system will now be fixed by the $f_1\qty(S,q)$ loop function. In this function there will be a dispersion relation with total momentum $P=q+k$, with $k$ the quark momentum (integrated up to $\Lambda_f$) and $q$ the external meson momentum (integrated up to $\Lambda_b$). It is clear that the maximum momentum in the system will be the sum $P_\mathrm{max}=\Lambda_f+\Lambda_b=\qty( 1 + \alpha )\Lambda_f$. If one considers that the NJL model is valid up to a momentum scale of $P_\mathrm{max}=1$ GeV, then $\alpha$ is limited by this energy scale for a given $\Lambda_f$. Hence, we will consider parametrizations where the ratio $\alpha$, yields a maximum momentum scale of the order of 1 GeV. More details will be given in Sec. \ref{results}.

\subsection{The one-meson-loop gap equation at finite temperature}

Calculating explicitly the functional derivatives in Eq.  (\ref{eq:gap_equation_definition}), one can arrive at the one-meson-loop gap equation, first derived in \cite{Nikolov:1996jj}:
\begin{align}
\Sigma_q \qty(S) + \Sigma_\sigma \qty(S) + \Sigma_\pi \qty(S) = 0.
\end{align}
The first term is the usual one-quark loop contribution while the remaining correspond to the $\sigma$ and $\pi$ one-meson-loop contributions to the gap equation. Each contribution is explicitly given by:
\begin{align}
\Sigma_q \qty(S) 
& = \frac{1}{g_s} \qty( S - m ) - 4 N_c N_f S  f_0\qty(S),
\label{gap_quark}
\\
\Sigma_\sigma \qty(S) 
& = 
2 N_c N_f S 
\qty{
4 I_{1\sigma} \qty(S) + 2 f_1\qty(S,0) I_\sigma \qty(S) + I_{2\sigma} \qty(S)
} , 
\label{gap_sigma}
\\
\Sigma_\pi \qty(S) 
& = 
6 N_c N_f S 
\qty{ 2 f_1\qty(S,0) I_\pi \qty(S) + I_{2\pi} \qty(S) } ,
\label{gap_pion}
\end{align}
where $I_M \qty(S)$, $I_{1M} \qty(S)$, and $I_{2M} \qty(S)$, with $M=\qty{ \sigma, \pi }$, are defined as:
\begin{align}
I_M \qty(S) & = 
\int\frac{ \dd[4]{q} }{\qty(2\pi)^4} 
\tilde{ \Delta }_M (S,q) ,
\label{def:IM}
\\
I_{1M} \qty(S) & = 
\int\frac{ \dd[4]{q} }{\qty(2\pi)^4} 
f_1\qty(S,q) 
\tilde{ \Delta }_M (S,q) ,
\label{def:I1M}
\\
I_{2M} \qty(S) & = 
- 2 \int\frac{ \dd[4]{q} }{\qty(2\pi)^4} 
\qty( q^2 + 4S^2 )f_2\qty(S,q) 
\tilde{ \Delta }_M (S,q)  .
\label{def:I2M}
\end{align}
The function $f_2\qty(S,q)$ can be written as a derivative of the $f_1\qty(S,q)$ loop function with respect to $S^2$. In the chiral limit, the one-loop corrections, $\Sigma_\sigma$ and $\Sigma_\pi$, are explicitly suppressed by an overall $N_c$ factor, due to the extra $N_c$ factor in the meson-loop propagator, $\tilde{ \Delta }_M (S,q)$, meaning that these terms are of $\mathcal{O}(N_c^0)$ \cite{Nikolov:1996jj,Florkowski:1996wf}.

In the meson loop corrections terms present in the gap equation (Eqs.  (\ref{gap_sigma}) and (\ref{gap_pion})), one is integrating over the meson four momentum $q$ i.e., summing over all kinematic meson fluctuations that can contribute to the system.

\begin{figure}[ht]
\centering
\includegraphics[width=0.4\textwidth]{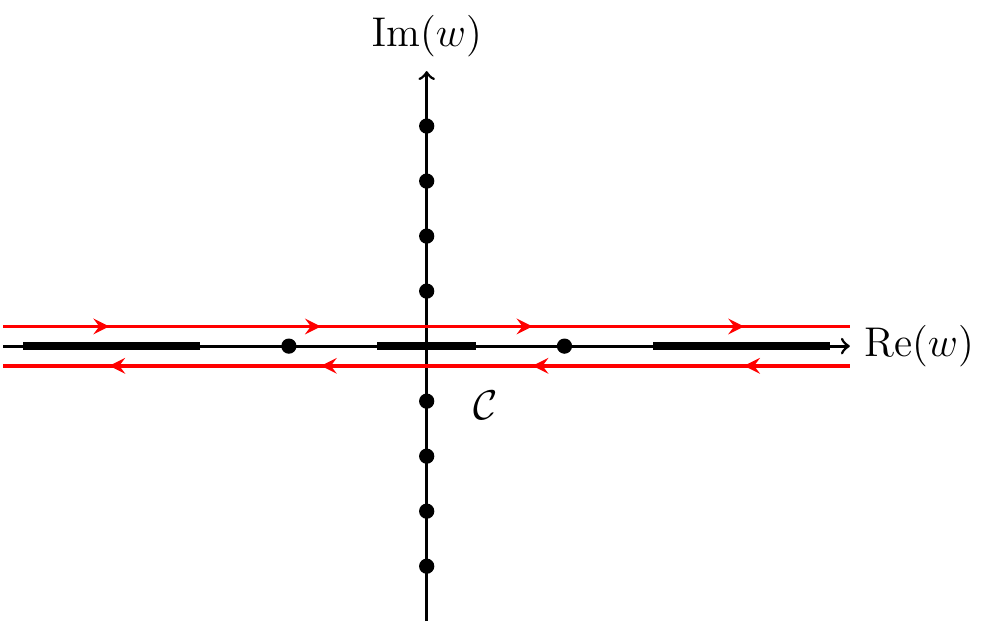}
\caption{\small Contour used to calculate the meson-loop contributions to the gap equation. The dots in the vertical axis are poles that represent the bosonic Matsubara frequencies. The poles and branch cuts on the horizontal axis are due to the analytical structure of the $f_1 \qty(S,q)$ loop function.}
\label{fig:contour2}
\end{figure}

At finite temperature, the meson-loop contributions can be calculated following the usual Matsubara sum technique and the vacuum can be calculated by taking the $T \to 0$ limit. These infinite sums over residues, of a previous singular integrand, can be transformed into a contour integration in the complex plane which avoid poles located at the Matsubara frequencies. However, the available contours in the complex plane are constrained by the analytical structure of the integrand. In this calculation, the meson propagator, more specifically the loop function $f_1 \qty(S,q)$, imposes restrictions on the possible contours in the complex plane. Hence, due to the analytic properties of such a function, the Matsubara sum will be transformed into a contour integration as suggested in Fig.  \ref{fig:contour2} (see Ref. \cite{HUFNER1994225}).

Each one-meson-loop term in the gap equation can be brought to a form of a contour integration of the function $h\qty(w)$ of a complex variable, $w$. The integral over the closed contour $\mathcal{C}$ in the complex plane (see Fig.  \ref{fig:contour2}), of the complex function $h\qty(w)$ can be written as:
\begin{align}
I  =
\oint_{\mathcal{C}} \frac{ \dd{w} }{ 2 \pi i } h\qty(w)  .
\label{eq:general_omega_integral_C2}
\end{align}
Defining the real part of $w$ as $\omega$, we can define the real and imaginary parts of the function $h\qty(w)$, near the real axis (small $\epsilon > 0$), by writing:
\begin{align}
h\qty(\omega \pm i \epsilon) = 
\Re \qty[ h\qty(\omega)  ] \pm i \Im \qty[ h\qty(\omega)  ] .
\end{align} 
The integration in Eq.  (\ref{eq:general_omega_integral_C2}), can then be written as an integration around the real axis as:
\begin{align*}
I
& =
\frac{1}{\pi}
\int_{-\infty}^{+\infty} \dd{\omega}
\Im \qty[ h\qty(\omega)  ]
\numberthis
\label{eq:general_contour_integration}
\end{align*}
Only the imaginary part of the function under the original contour integration, $\Im \qty[ h\qty(\omega)  ]$, will contribute to the result.

\begin{figure}[ht]
\centering
\includegraphics[width=0.4\textwidth]{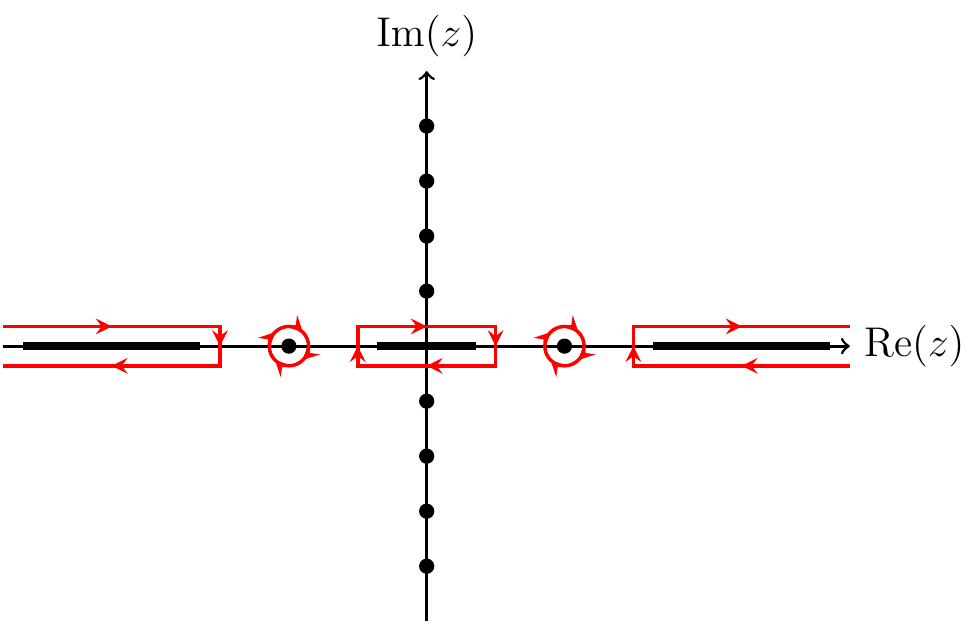} 
\caption{\small Definition of the collective meson mode (pole) and the noncollective meson mode (branch cut) terms in the meson-loop corrections.} 
\label{fig:contour3}
\end{figure}

In our framework, to calculate the meson contributions for a given meson channel $M$, two distinct contributions will be considered, the collective and noncollective modes (see Refs. \cite{YAMAZAKI201319,YAMAZAKI2014237}). This separation is depicted in Fig.  \ref{fig:contour3}, where the first comes from the isolated pole in the complex plane while the latter, from the branch cuts. 

In the chiral limit, Eq.  (\ref{eq:tilde_meson_propagator_def}) can be viewed as the propagator of a meson with effective mass $4S^2 \delta_{M \sigma}$ and a wave function renormalization proportional to $f_1\qty(S,q)$. The function $f_1\qty(S,q)$ amounts for the internal interacting quark substructure of the collective meson excitation. For convenience, we define the \textit{collective} meson propagator in Eq. (\ref{eq:tilde_meson_propagator_def}) as:
\begin{align}
k_M \qty(S,\vec{q}, q_0 ) =
\frac{ 1 }{\qty(q^2 + 4S^2 \delta_{M\sigma} )} = 
\frac{ 1 }{ q_0^2 + E^2_M \qty(S,\vec{q}) } ,
\label{eq:collective_meson_prop}
\end{align}
where the dispersion relation $E_M^2 = { \vec{q}^2 + 4S^2\delta_{M\sigma} }$, is a real quantity.

The collective mode contributions will be calculated by considering that only the \textit{collective} meson propagator, $k_M \qty(S,\vec{q},-iw)$, has a nonvanishing imaginary part and $f_1 \qty(S,\vec{q},-iw)$ is a real quantity. These pole terms will appear as delta functions and  will correspond to excitations of the underlying quark system with a precise dispersion relation. The noncollective modes come from the branch cuts, corresponding to the kinematic region where the imaginary part of $f_1 \qty(S,\vec{q},-iw)$  is nonzero and the \textit{collective} meson propagator, $k_M \qty(S,\vec{q},-iw)$ is a real quantity. The analytic continuation of the functions $f_1\qty(S, \vec{q}, q_0)$ to $f_1\qty(S, \vec{q}, -iq_0)$ and $k_M\qty(S, \vec{q}, q_0)$ to $k_M\qty(S, \vec{q}, -iq_0)$ have been defined as $F \qty(S, \vec{q}, \omega)$ and $K_M\qty(S, \vec{q}, \omega)$, respectively, for real $\omega$ (see Appendixes \ref{appendix_f1_loop_function} and \ref{appendix_IM}).

As pointed out by K. Yamazaki \textit{et al.} in Refs. \cite{YAMAZAKI201319,YAMAZAKI2014237}, when chiral symmetry is not explicitly broken at the Lagrangian level, i.e., when $m=0$, these contributions are easily separated.  When including the quark current mass however, these contributions get mixed and the separation must be done with care.

As the temperature increases, one expects chiral symmetry to get restored. This means that the absolute value of the quark condensate decreases, as well as the value for the expectation value of the scalar field, $S$. This implies that, both the position of the meson propagator pole in the complex plane, as well as the onset of the branch cuts, can change with the temperature and $S$. 

In the MF calculation of meson masses and decays, one can define the Mott temperature at which the mass of a given meson channel, is smaller then the sum of the constituent mass of its composing quarks (for a detailed discussion in the two flavor NJL model, see \cite{Hansen:2006ee}). At this point the decay width of such a meson channel is nonzero and the previous quarks bound state, becomes a resonance.  In the present paper, this corresponds to the meson pole reaching the branch cut. At this point, both the \textit{collective} meson propagator, $k_M \qty(S,\vec{q},-iw)$ and the loop function $f_1 \qty(S,\vec{q},-iw)$ have nonvanishing imaginary parts. To calculate exactly such contributions, one should use a keyhole contour, avoiding both the pole as well as the branch cut singularity. However, that would introduce in the calculation a mixture between the imaginary contribution coming from the pole with the one coming from the cut, making it very difficult to clearly separate both contributions. To avoid this, in the present framework, for a given kinematic contribution where the pole lies on top of the branch cut, only the noncollective mode will be calculated.

A collective meson mode exists, if there is an $\omega_\P$ value,  in-between the branch cuts, where Eq.  (\ref{eq:tilde_meson_propagator_def}) is zero. This condition can be written as:
\begin{align}
-\omega_\P^2 + E_M^2\qty(S,q) + \frac{ \tilde{m} }{ \Re \qty[ F \qty(S,\vec{q}, \omega_\P) ]  } = 0 ,
\label{eq:pole_condition}
\end{align} 
where,
\begin{align}
\tilde{m} =  \frac{m}{ 2 g_s N_c N_f  S} .
\end{align}
Analyzing $\Im \qty[ F \qty(S,\vec{q}, \omega) ]$, one can recognize that the region in-between cuts is given by $[ \sqrt{ \qty(\Lambda_f+q)^2 + S^2 } - \sqrt{ \Lambda_f^2 + S^2 }, E_\sigma ]$. One can also observe that the real part of $F \qty(S, \vec{q}, \omega)$ is always greater then zero in the $\omega-$region in-between the branch cuts. Thus, considering a finite current quark mass, Eq.  (\ref{eq:pole_condition}) only has a zero for the pion meson mode. This means that excitations with the same quantum numbers as the $\sigma$ field will not have collective mode contributions, only noncollective ones.

In the following, the integrations defined in Eqs.  (\ref{def:IM}), (\ref{def:I1M}) and (\ref{def:I2M}), will be separated in the collective and noncollective contributions.

Consider the contribution $I_M \qty(S)$, for a given meson channel $M=\{ \sigma, \pi \}$, given in Eq.  (\ref{def:IM}) (for more details on this calculation, see Sec. \ref{appendix_IM}). As discussed earlier, it can be divided in the collective and noncollective contributions, i.e., the pole $\mathcal{P}_M \qty(S)$ and a branch cut, $\mathcal{B}_M \qty(S)$ terms. This separation can be written as:
\begin{align}
I_M \qty(S) & = \mathcal{P}_M \qty(S) + \mathcal{B}_M \qty(S).
\label{def:IM_pole_cut_decomposition}
\end{align}
The first term is the contribution coming from the collective modes. It can be calculated, as already stated, by considering that near the real axis, the loop function $f_1\qty(S,\vec{q},-i w)$ is purely real and $k_M^{-1} \qty(S,\vec{q},-i w)$ has both a real and an imaginary part. It can be calculated to yield:
\begin{align*}
\mathcal{P}_M \qty(S) 
& = 
\frac{1}{ 4 N_c N_f }
\int_{ \vec{q} }
\frac{ \coth \qty( \nicefrac{ \beta \omega_\P }{ 2 } ) }{ \Re \qty[ F \qty(S,\vec{q}, \omega_\P ) ] }
\frac{ \abs{ \partial_\omega \chi_\P \qty(S,\vec{q},\omega ) }_{\omega_\P}^{-1} }{ \tilde{E}_M (S,\vec{q}, \omega_\P ) }   .
\numberthis
\end{align*}
Here, the collective mode dispersion relation $\tilde{E}_M^2 (S,\vec{q},\omega)$ and the function $\chi_\P \qty(S,\vec{q},\omega)$, are defined as:
\begin{align}
\tilde{E}_M^2 (S,\vec{q},\omega) & = 
E_M^2(S,\vec{q}) +
\frac{ \tilde{m} }{ \Re \qty[ F \qty(S,\vec{q}, \omega) ]  } ,
\\
\chi_\P \qty(S,\vec{q},\omega) & = \omega - \tilde{E}_M (S,\vec{q},\omega) ,
\end{align}
while $\omega_\P = \omega_\P \qty(S, \vec{q})$ is the location of the pole on the real line of the $\omega-$complex plane. It can be calculated as a solution of
\begin{align}
\chi_\P \qty(S,\vec{q},\omega_\P)=0 .
\label{def:pole_location_eq}
\end{align}

Now, one of the difficulties of including composite meson fluctuations in the calculation becomes evident. The pole location $\omega_\P$, from which one calculates the collective mode dispersion relation $\tilde{E}_M^2 (S,\vec{q},\omega_\P)$, depends on the Hartree mass ($S$), on the meson 3-momentum ($\vec{q}$) and implicitly on the temperature ($T$), through $\Re \qty[ F \qty(S,\vec{q}, \omega) ]$, which is related to the quark substructure of the collective mode. 

From this, one can see that the pole contribution, does not simply correspond to a integration over the meson fluctuation momentum with a fixed collective meson mass. When integrating over the meson momentum, a certain value of Hartree mass and temperature are fixed and the pole location, for a single value of $\vec{q}$, is calculated self-consistently. We highlight that, in our calculation, the pole contributions are only nonzero if $\omega_\P$ exists in between the cuts.

The second term, $\mathcal{B}_M \qty(S)$, can be calculated by considering that, near the real axis, $k_M^{-1} \qty(S,\vec{q},-i w)$ is real while $f_1\qty(S,\vec{q},-i w)$ is complex. One can write:
\begin{align}
\mathcal{B}_M \qty(S) 
& = 
\frac{1}{ 2 \pi N_c N_f }
\int_{ \vec{q} }
\int_{ 0 }^{+\infty} \dd{\omega}
\frac{ \coth \qty( \nicefrac{ \beta \omega }{ 2 } ) }{ -\omega^2 + { {E} }^2_M \qty(S,\vec{q}) }
\frac{ - \Im \qty[ F \qty(S,\vec{q}, \omega ) ]  }{ \Re \qty[ G \qty(S,\vec{q}, \omega ) ]^2 + \Im \qty[ F \qty(S,\vec{q}, \omega ) ]^2 }
.
\end{align}
The function $\Im \qty[ F \qty(S,\vec{q}, \omega ) ]$ have an Heaviside step function, which restricts the integration to the branch cuts in Fig.  \ref{fig:contour3}. The function $\Re \qty[ G \qty(S,\vec{q}, \omega ) ]$, is defined as:
\begin{align}
\Re \qty[ G \qty(S,\vec{q}, \omega ) ] = 
\Re \qty[ F \qty(S,\vec{q}, \omega ) ] + \tilde{m} K_M \qty(S,\vec{q},\omega ) .
\label{eq:def_ReG}
\end{align}

%%%%%%%%%%%%%%%%%%%%%%%%%%%%%%%%%%%%%%%%%%%%%%%%%%%%%%%%%%%%%%%%%%%%%%%%%%%%%%%%%%%%%%%%%%%%%%%%
%%%%%%%%%%%%%%%%%%%%%%%%%%%%%%%%%%%%%%%%%%%%%%%%%%%%%%%%%%%%%%%%%%%%%%%%%%%%%%%%%%%%%%%%%%%%%%%%

The integral in Eq.  (\ref{def:I1M}), only appears in the $\sigma$ gap equation. Considering a finite quark current mass $m$,  only the branch cut contribution will be nonzero, $I_{1\sigma} \qty(S) = \mathcal{B}_{1\sigma} \qty(S)$ since, as previously stated, the $\sigma$ mode does not have a pole. One can write this term as (see Sec. \ref{appendix_I1M} for more details on this derivation):
\begin{align}
\mathcal{B}_{1\sigma} \qty(S)
& = 
\frac{ \tilde{m} }{ 2 \pi N_c N_f }
\int_{ \vec{q} }
\int_{ 0 }^{+\infty} \dd{\omega}
\frac{ \coth \qty( \nicefrac{ \beta \omega }{ 2 } ) }{ -\omega^2 + { {E} }^2_\sigma \qty(S,\vec{q}) }
\frac{ K_\sigma \qty(S,\vec{q},\omega ) \Im \qty[ F \qty(S,\vec{q}, \omega ) ] }{ \Re \qty[ G \qty(S,\vec{q}, \omega ) ]^2 + \Im \qty[ F \qty(S,\vec{q}, \omega ) ]^2  } .
\end{align}
It is clear that this contribution vanishes in the chiral limit, due to the overall factor $\tilde{m}$.

%%%%%%%%%%%%%%%%%%%%%%%%%%%%%%%%%%%%%%%%%%%%%%%%%%%%%%%%%%%%%%%%%%%%%%%%%%%%%%%%%%%%%%%%%%%%%%%%
%%%%%%%%%%%%%%%%%%%%%%%%%%%%%%%%%%%%%%%%%%%%%%%%%%%%%%%%%%%%%%%%%%%%%%%%%%%%%%%%%%%%%%%%%%%%%%%%

The last integration that needs attention, is given by Eq.  (\ref{def:I2M}) (for more details see Sec. \ref{appendix_I2M}). It will have contributions coming both from the collective and noncollective modes:
\begin{align}
I_{2M} \qty(S) & = \mathcal{P}_{2M} \qty(S) + \mathcal{B}_{2M} \qty(S).
\label{def:I2M_P2M_plus_B2M}
\end{align}
To simplify the calculations one can write the integrand in terms of the $f_1 \qty(S,q)$ loop function using the identity:
\begin{align}
f_2 \qty(S,q) =
- \frac{1}{2}
\pdv{ \xi^2 }
f_1 \qty(\xi,q)
_{\xi=S} .
\label{eq:f2_f1_relation}
\end{align}
This will remove double poles that would otherwise appear when using the Matsubara sum technique.

Repeating the same process i.e., consider that $f_1\qty(S,\vec{q},-i w)$ is purely real and $k_M^{-1} \qty(S,\vec{q},-i w)$ is complex, near the real axis, after some calculations, one can arrive at:
\begin{align*}
\mathcal{P}_{2M} \qty(S)
& =
-
\frac{ \tilde{m} }{ 4 N_c N_f }
\int_{ \vec{q} }
\frac
{ \coth \qty( \nicefrac{ \beta \omega_\P }{ 2 } ) }
{ \tilde{E}_M (S,\vec{q},\omega_\P) }
\frac
{ \partial_{S^2} \Re \qty[ F \qty(S,\vec{q}, \omega_\P) ] }
{ \Re \qty[ F \qty(S,\vec{q}, \omega_\P) ]^2  }
\abs{ \partial_\omega \chi_\P \qty(S,\vec{q},\omega ) }_{\omega_\P}^{-1}  .
\numberthis
\end{align*}

The noncollective contribution to $I_{2M} \qty(S)$ can be calculated as before, near the real axis, the branch cut term is:
\begin{align}
\mathcal{B}_{2M} \qty(S)
& =
\frac{1}{ 2 \pi N_c N_f }
\int_{ \vec{q} }
\int_{0}^{+\infty} \dd{\omega}
\frac{ \coth \qty( \nicefrac{ \beta \omega }{ 2 } ) }{1 + A \qty(S,\vec{q}, \omega)^2} 
\partial_{\xi^2} A \qty(\xi,\vec{q}, \omega)_{\xi=S}  .
\end{align}
Here, the function $A \qty(S,\vec{q}, \omega)$ is defined as
\begin{align}
A \qty(\xi,\vec{q}, \omega) & = 
\frac
{ \Im \qty[F \qty(\xi,\vec{q}, \omega)] }
{ \Re \qty[ F \qty(\xi,\vec{q}, \omega ) ] + \tilde{m} K_M \qty(S,\vec{q},\omega ) }  .
\end{align}

\section{Results}
\label{results}

In this section we present our results and discuss the influence of the one-meson-loop terms, separated in collective and noncollective contributions, on the quark condensate in the vacuum and at finite temperature. We also study the effect of including only the collective and noncollective contributions in the restoration of chiral symmetry with increasing temperature.

Here, we point out that, concerning the numerical calculations, the inclusion of the one-meson-loop terms is completely self-consistent: upon solving the gap equation for a given parametrization, for each value of Hatree mass, $S$, and temperature, $T$, one has to numerically check the existence of the collective modes and their influence on the noncollective modes.

\subsection{Vacuum}

To start our study, we find a parameter set which, at the MF level, reproduces the value of the quark condensate obtained by two-flavor lattice QCD \cite{Cichy:2013gja}, $\expval{\bar{\ell} \ell}^{\nicefrac{1}{3}}=-256$ MeV, the pion mass, $m_{\pi}=135$ MeV and the pion decay constant, $f_{\pi}=93$ MeV. This parameter set is displayed in Table \ref{MF_parameters}.

\begin{center}
\begin{table}[h!]
\begin{tabular}{|c|c|c||c|}
\hline
$\Lambda_f \qty[\textrm{MeV}]$ & $m \qty[\textrm{MeV}]$ & ${g_s}\Lambda_f^2/2 $ & $S \qty[\textrm{MeV}]$ \\
\hline
690.3 & 4.72 & 2.014 & 288.4 \\
\hline
\end{tabular}
\caption{Mean field parameter set and MF quark mass, $S$, in the vacuum.}
\label{MF_parameters}
\end{table}
\end{center}

\begin{figure}[h!]
\centering
\begin{subfigure}{.5\textwidth}
\centering
\includegraphics[width=1.\linewidth]
{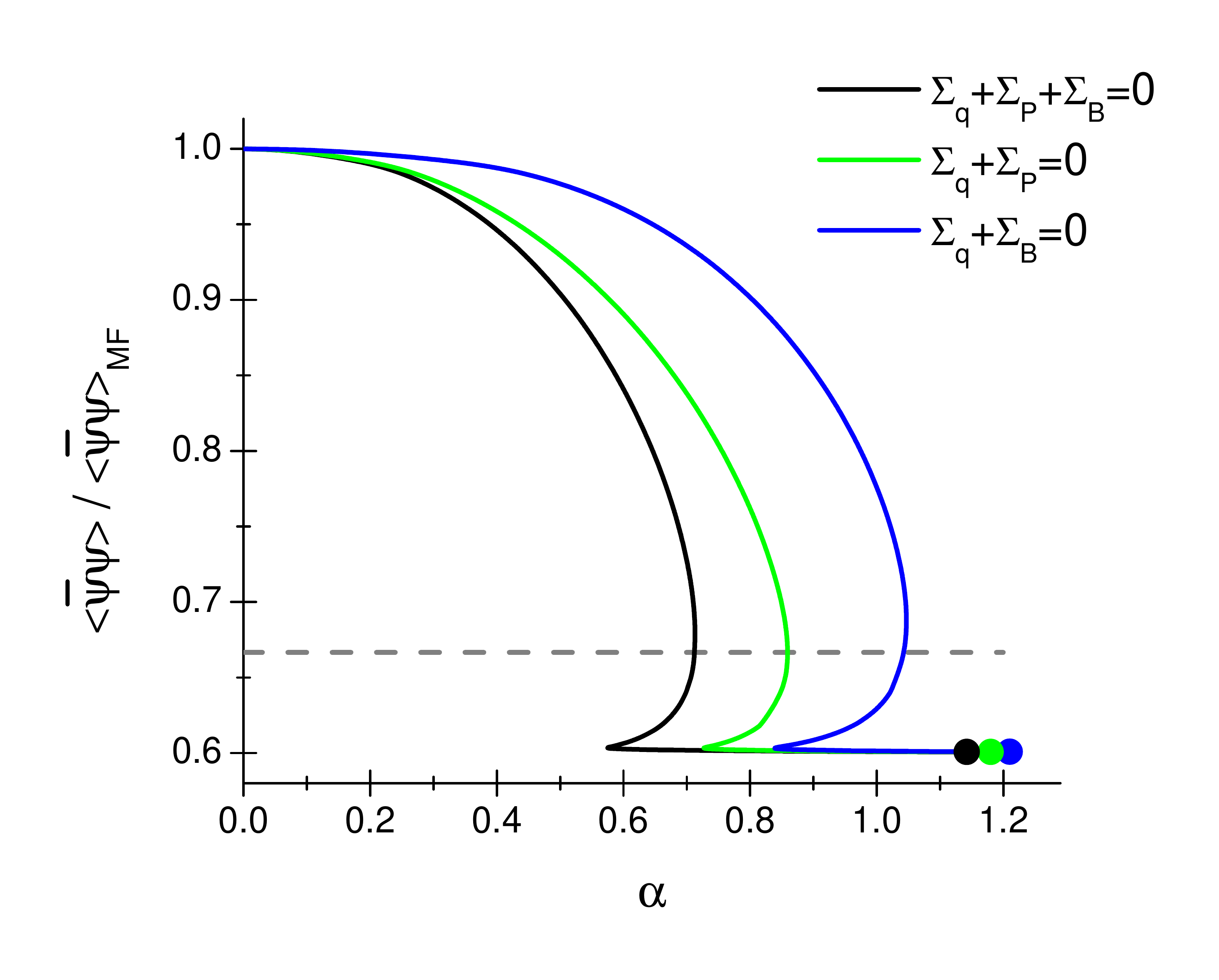}
%\caption{A subfigure}
\label{fig:sub1}
\end{subfigure}%
\begin{subfigure}{.5\textwidth}
\centering
\includegraphics[width=1.\linewidth]
{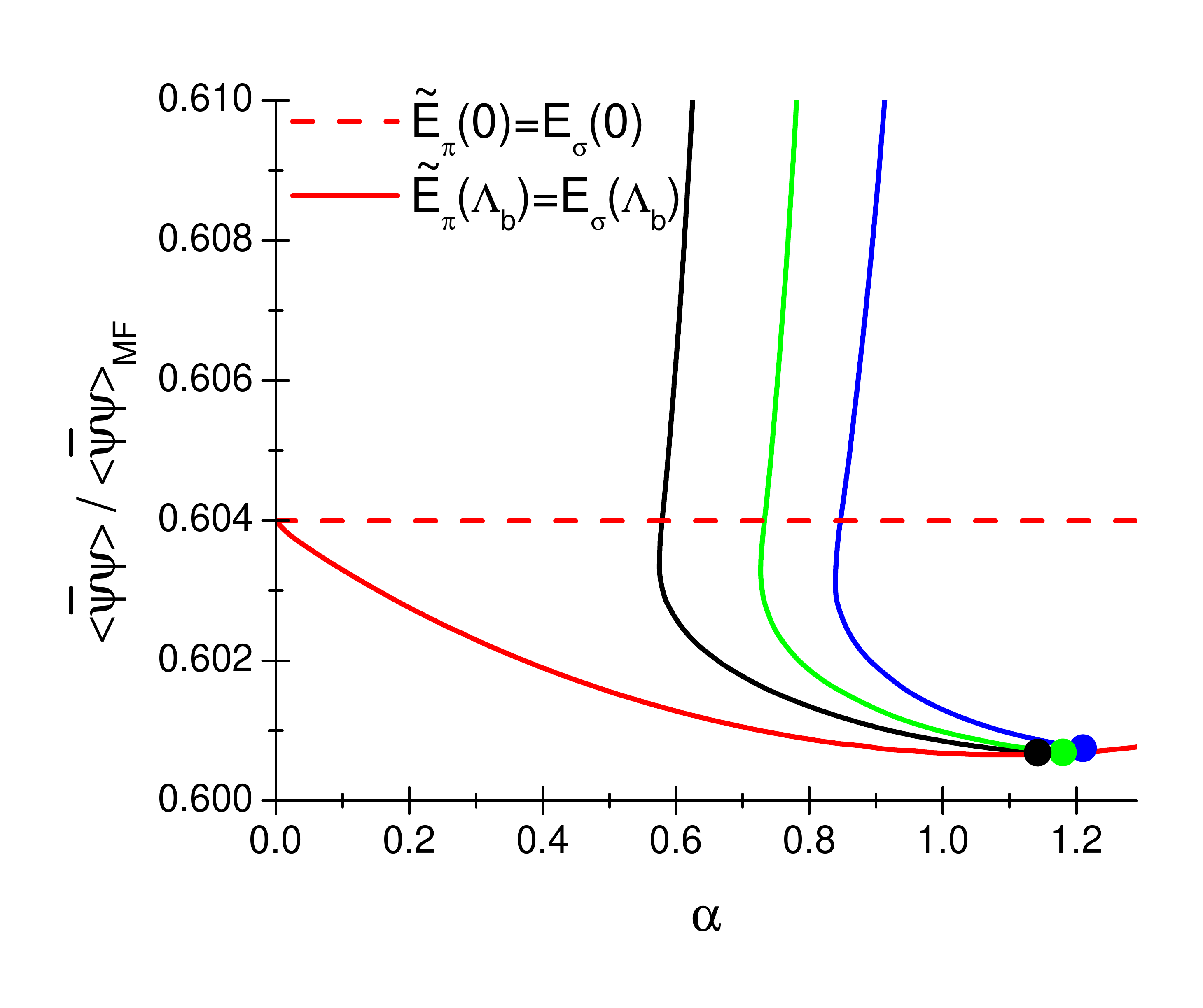}
%\caption{A subfigure}
\label{fig:sub2}
\end{subfigure}
\caption{\small Ratio between the MF vacuum quark condensate  and the one-meson-loop vacuum quark condensate, as a function of the ratio $\alpha=\Lambda_b/\Lambda_f$. The green line is the result of solving the gap equation with the collective contributions, the blue line with the noncollective contributions and the black line is the complete calculation. The gray dashed line in the left panel, corresponds to an $1/N_c$-reduction of the MF vacuum quark condensate. The red-dashed and red-full lines in the right panel, correspond to the Hartree mass points where the $\pi$ collective mode reaches the branch cut, with $q=0$ and  $q=\Lambda_b$, respectively. }
\label{vacuum_results}
\end{figure}

To study the effect of the inclusion of meson-loop corrections in the vacuum condensate, we use the aforementioned MF parameter set and increase the value of $\alpha$, the ratio between the boson and fermion cutoff, $\alpha=\Lambda_b/\Lambda_f$, from zero (MF calculation) to a finite value. The results of such calculation can be seen in Fig.  \ref{vacuum_results}. Three different scenarios were considered: 
\begin{itemize}

\item Quark loop and the collective modes ,  $\Sigma_\mathrm{q}(S) +\Sigma_\mathrm{P}(S)=0$;

\item Quark loop and the noncollective modes, $\Sigma_\mathrm{q}(S)+\Sigma_\mathrm{B}(S)=0$;

\item Quark loop and collective and noncollective modes,  $\Sigma_\mathrm{q}(S) +\Sigma_\mathrm{P}(S)+\Sigma_\mathrm{B}(S)=0$.

\end{itemize}

Setting the boson cutoff to a nonzero value is equivalent to include the one-meson-loop correction terms. As one can see in the left panel of Fig.  \ref{vacuum_results}, by solving the gap equation with increasing $\alpha$, the value of the quark condensate decreases. For reference, the gray dashed line in the left panel of Fig.  \ref{vacuum_results}, corresponds to an $1/N_c$-reduction of the MF vacuum quark condensate. This decreasing behavior is expected since the inclusion of bosonic degrees of freedom is known to restore chiral symmetry. The decreasing of the quark condensate with increasing $\alpha$ happens until a point where, to further decrease the quark condensate, the boson cutoff has also to decrease. This behavior of decreasing quark condensate with decreasing $\alpha$, continues up to the point where the pion collective mode with zero momentum reaches the branch cut i.e., $\tilde{E}_\pi\qty(0)=E_\sigma\qty(0)$. This can been seen more clearly in the right panel of of Fig.  \ref{vacuum_results}. After this point (red-dashed line in the right panel of Fig.  \ref{vacuum_results}) a smaller number of momentum modes will contribute to the collective modes and the quark condensate cannot decrease again with increasing $\alpha$. When the highest momentum mode, with $q=\Lambda_b$, reaches the branch cut i.e., $\tilde{E}_\pi\qty(\Lambda_b)=E_\sigma\qty(\Lambda_b)$, the collective modes do not contribute any more to the calculation (full red line in the right panel of Fig.  \ref{vacuum_results}). At this point, no more solutions can be found for the gap equations. These points are represented in the right panel of Fig.  \ref{vacuum_results} by the respective colored dots.

\subsection{Finite temperature}

In this section we solve the gap equation at finite temperature for different sets of parameters that include one-meson-loop corrections and compare the results with the usual MF calculation.

To solve the gap equation at finite temperature it is necessary to evaluate the $q \to 0$ limit of the $ {f_1 (S,q) } $ loop function i.e., $f_1\qty(S,0)$ (see Eqs.  (\ref{gap_sigma}) and (\ref{gap_pion})). This operation implies two distinct limits, $q_0 \to 0$ and $\vec{q} \to 0$. After the extension of the discrete Matsubara frequencies to continuum values $q_0$, the function $ {f_1 (S,q) } $ is no longer analytic in the origin \cite{das1997finite}. This can easily be demonstrated by noticing that the limiting operations, $\vec{q} \to 0$ and $q_{0} \to 0$, do not commute i.e.,
\begin{align}
\lim_{\vec{q} \to 0} \lim_{q_{0} \to 0} f_1 \qty(S,\vec{q},q_0)
\neq
\lim_{q_{0} \to 0} \lim_{\vec{q} \to 0} f_1 \qty(S,\vec{q},q_0) .
\label{limiting_operations}
\end{align}
This is a consequence of the breaking of Lorentz symmetry by the heat bath. In fact, this feature is a well-know property of finite temperature field theory and the limiting operations in Eq.  (\ref{limiting_operations}) are related to two distinct approximations. The left-hand side order of limiting operations is known as the \textit{static} limit while, the one in the right-hand side, is known as the \textit{plasmon} limit. The analytical result for both limits is presented in Appendix \ref{appendix_f1_loop_function}. For more details see \cite{das1997finite}. We consider both the \textit{static} and \textit{plasmon} limits and compare both results in the calculation of the quark condensate as a function of temperature including collective and noncollective modes.

To study the finite temperature behavior of the quark condensate and restoration of chiral symmetry with the one-meson-loops contribution, a set of parameters has to be provided which include the boson cutoff. In order to do so, we fix the ratio between the boson and fermion cutoffs, $\alpha$, to different values and search for parametrizations which reproduce the same vacuum observables as in the MF case: the two flavor quark condensate, the pion mass and the pion decay constant given previously. We also search for parametrizations in the three scenarios presented earlier, considering the complete one-meson-loop gap equation, and considering the quark loop with the collective excitations or with the noncollective excitations. The obtained parameter sets are displayed in Table \ref{alpha_parameter_table}. 

To obtain the model parametrization, the pion mass and pion decay constant are calculated using the meson-loop pion propagator given in Eq.  (\ref{eq:tilde_meson_propagator_def}). We highlight that this is an approximation since the vacuum quantities are not calculated using the one-meson-loop pion propagator i.e., the second functional derivative of the one-loop effective action. This approximation only changes the parametrization of the model and does not modify the qualitative effects of including collective and noncollective modes on the quark condensate and on the restoration of chiral symmetry.

\begin{table}[ht!]
\begin{tabular}{|c|c|c|c|c|c||c|}
\hline
$\alpha$ & $\sum_\mathrm{P}$ & $\sum_\mathrm{B}$ & $\Lambda_f \qty[\textrm{MeV}]$ & $m \qty[\textrm{MeV}]$ & ${g_s}\Lambda_f^2/2 $ & $S \qty[\textrm{MeV}]$ \\ \hline
      & \cmark & \cmark & 690.9 & 4.72  & 2.015 & 288.1  \\ \cline{2-7} 
$0.1$ & \cmark & \xmark & 690.8 & 4.72  & 2.015 & 288.2  \\ \cline{2-7} 
      & \xmark & \cmark & 690.4 & 4.72  & 2.015 & 288.4  \\ \hline
      & \cmark & \cmark & 694.4 & 4.72  & 2.022 & 286.2  \\ \cline{2-7} 
$0.2$ & \cmark & \xmark & 693.7 & 4.72  & 2.021 & 286.5  \\ \cline{2-7} 
      & \xmark & \cmark & 691.0 & 4.72  & 2.016 & 288.1  \\ \hline
      & \cmark & \cmark & 702.2 & 4.72  & 2.038 & 282.1  \\ \cline{2-7} 
$0.3$ & \cmark & \xmark & 693.7 & 4.72  & 2.021 & 286.5  \\ \cline{2-7} 
      & \xmark & \cmark & 692.6 & 4.72  & 2.019 & 287.2  \\ \hline
      & \cmark & \cmark & 714.7 & 4.72  & 2.065 & 276.0  \\ \cline{2-7} 
$0.4$ & \cmark & \xmark & 709.2 & 4.72  & 2.053 & 278.6  \\ \cline{2-7} 
      & \xmark & \cmark & 695.7 & 4.72  & 2.025 & 285.5  \\ \hline
\end{tabular}
\caption{Parameter sets for different values of $\alpha$, considering three different scenarios: the complete calculation, considering only the quark sector and collective fluctuations and quark sector and noncollective fluctuations.}
\label{alpha_parameter_table}
\end{table}

\begin{figure}[ht!]
\centering
%%%%%%%
\begin{subfigure}[b]{0.475\textwidth}
\centering
\includegraphics[width=\textwidth]
{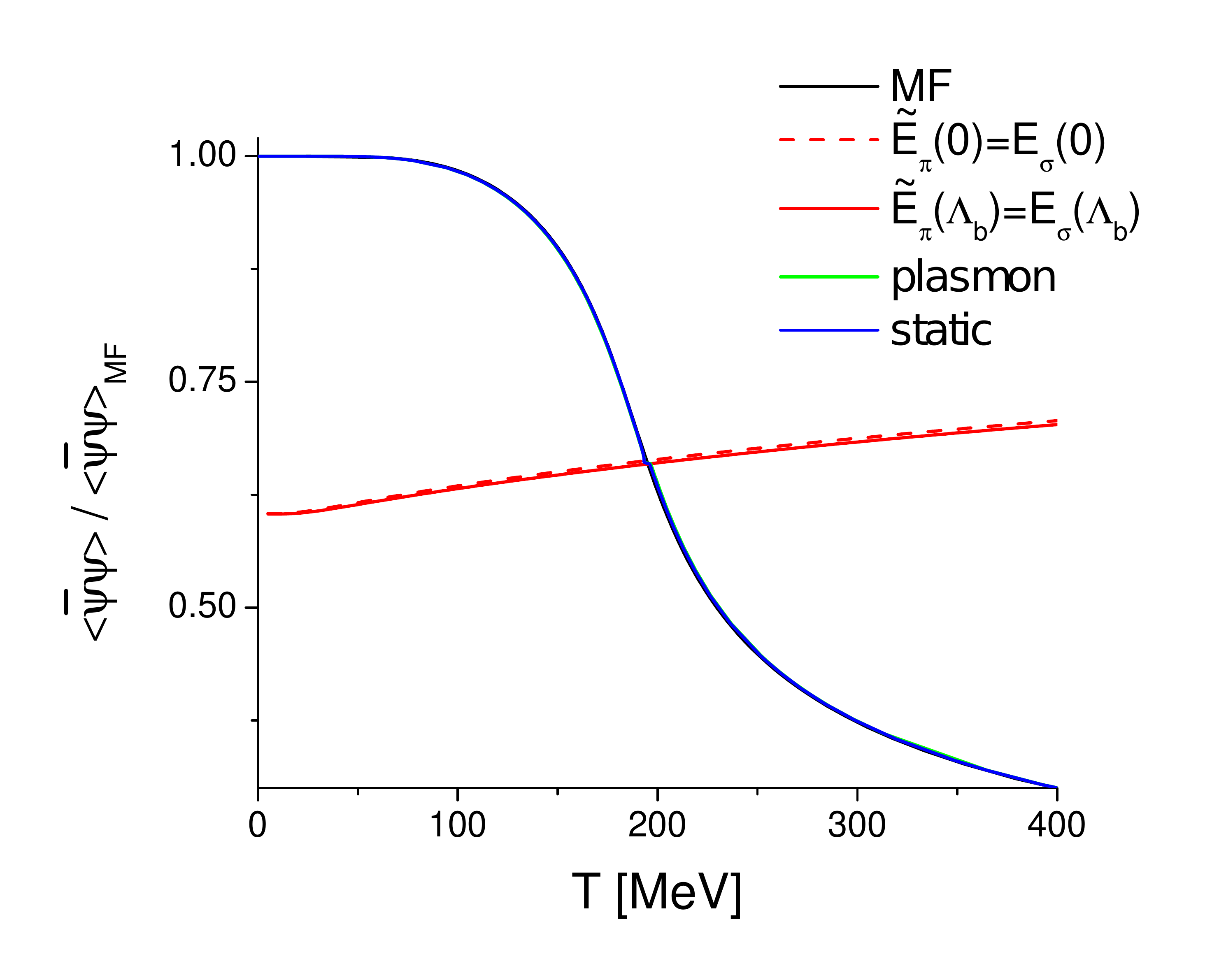}
\caption{$\alpha=0.1$}    
\label{fig:alpha_01}
\end{subfigure}
%%%%%%
\hfill
%%%%%%
\begin{subfigure}[b]{0.475\textwidth}  
\centering 
\includegraphics[width=\textwidth]
{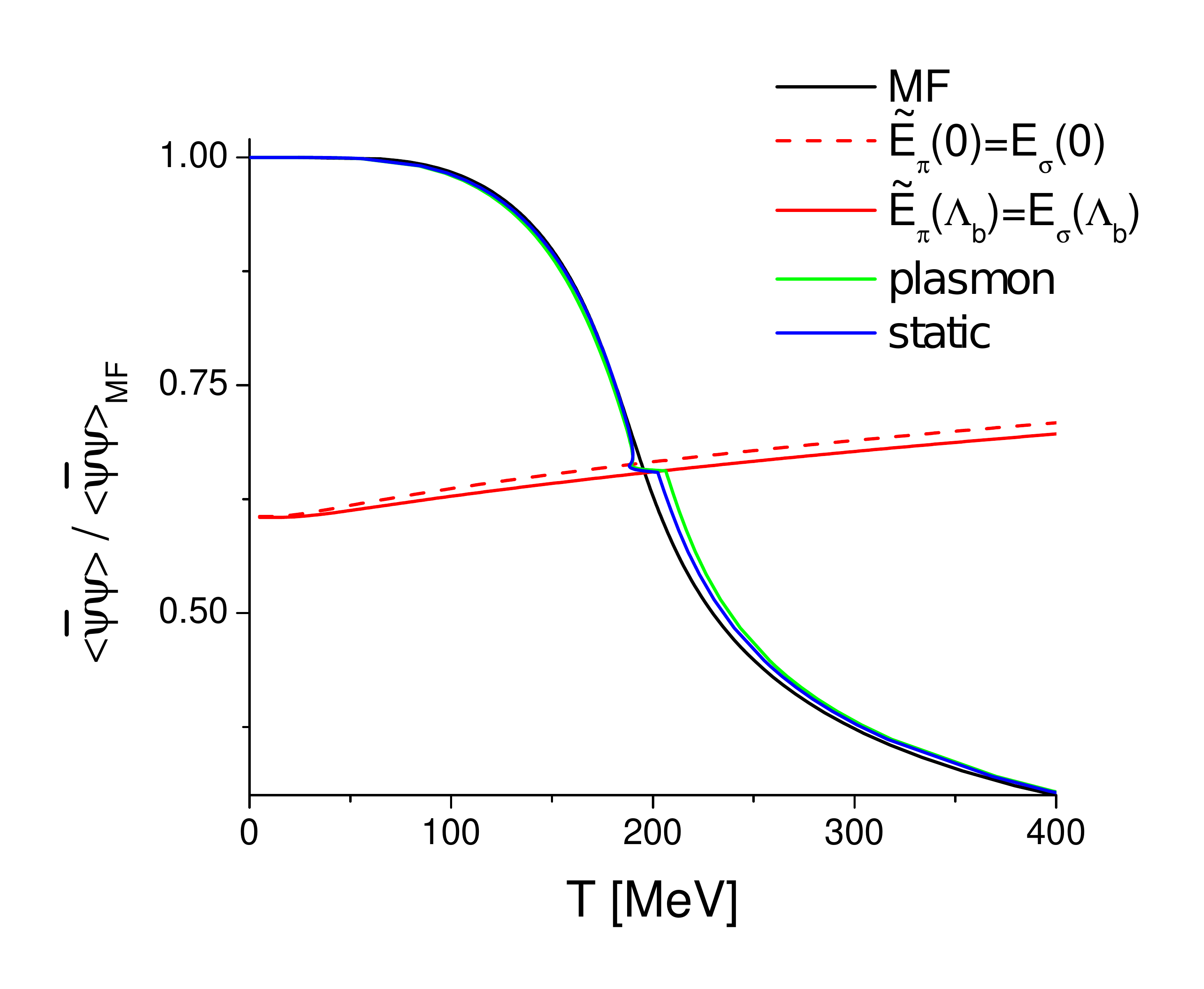}
\caption{$\alpha=0.2$}    
\label{fig:alpha_02}
\end{subfigure}
%%%%%%
%%%%%%
%%%%%%
\vskip\baselineskip
%%%%%%
%%%%%%
%%%%%%
\begin{subfigure}[b]{0.475\textwidth}   
\centering 
\includegraphics[width=\textwidth]
{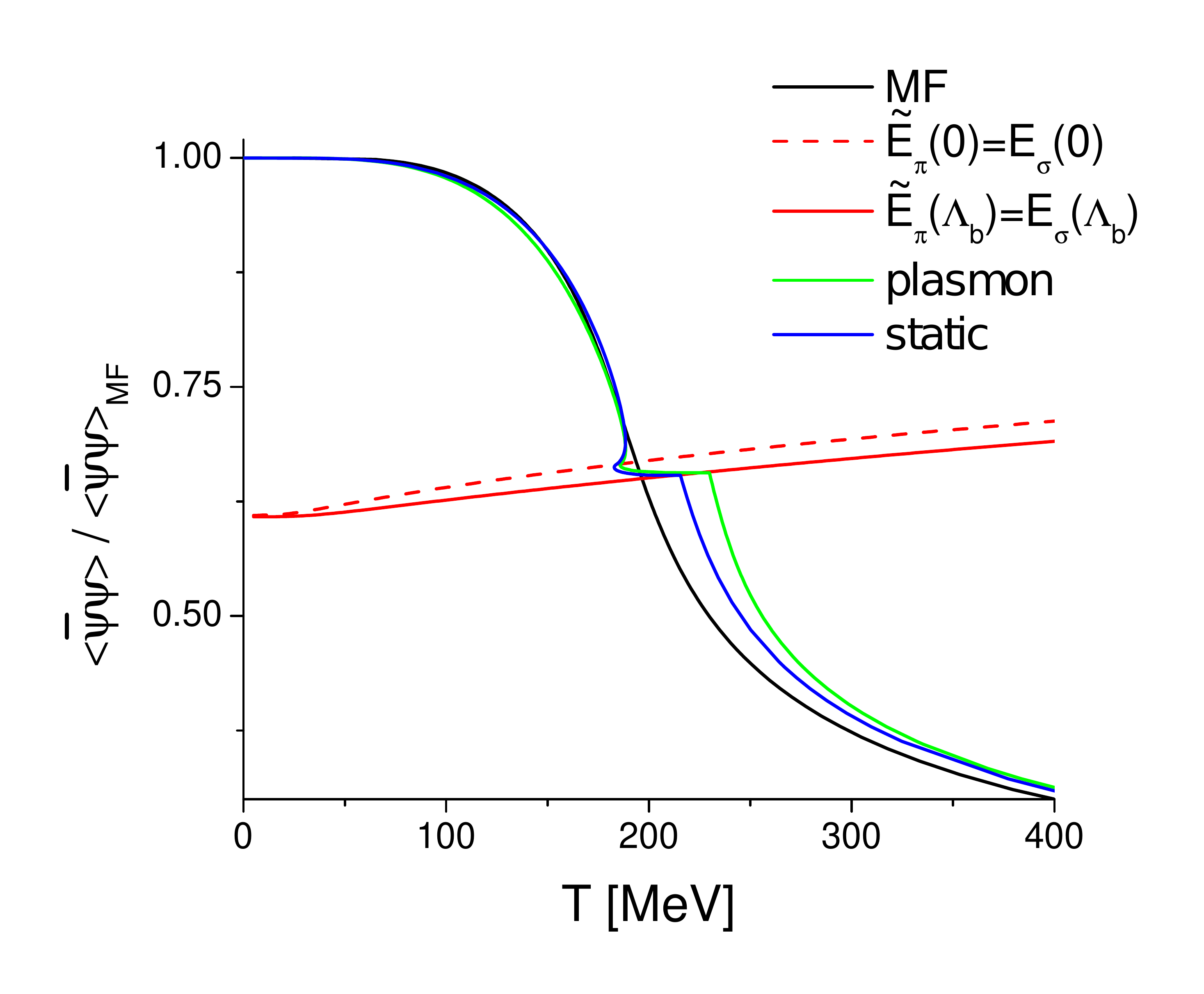}
\caption{$\alpha=0.3$}    
\label{fig:alpha_03}
\end{subfigure}
%%%%%%
\hfill
%%%%%%
\begin{subfigure}[b]{0.475\textwidth}   
\centering 
\includegraphics[width=\textwidth]
{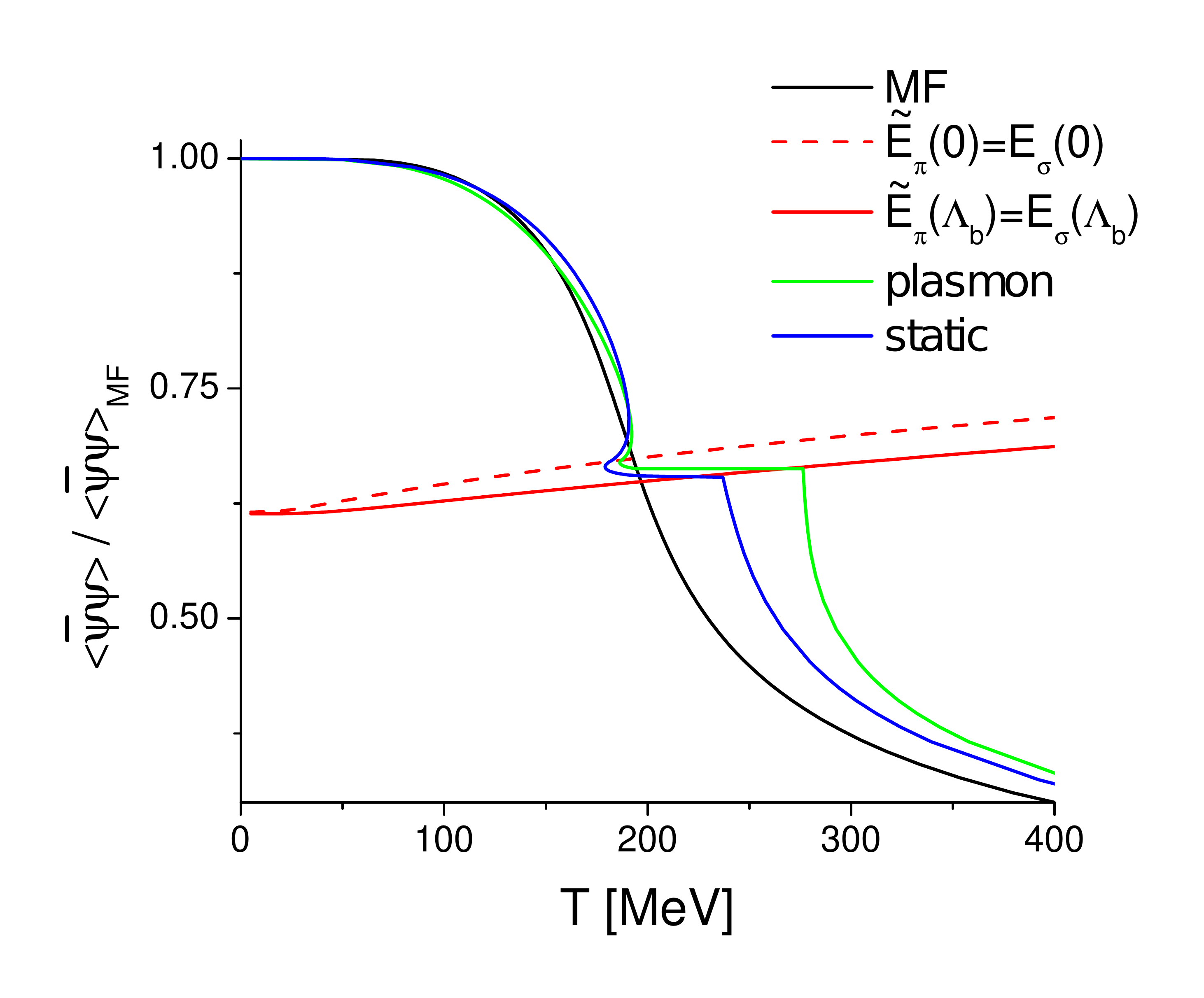}
\caption{$\alpha=0.4$}    
\label{fig:alpha_04}
\end{subfigure}
%%%%%%
%%%%%%
%%%%%%
\caption{\small Solution of the gap equation at finite temperature including collective and noncollective fluctuations. Each panel represents the solution for a given ratio between the boson and fermion cutoff, $\alpha = \Lambda_b/\Lambda_f$. Both the \textit{plasmon} and \textit{static} limits are presented as well as the collective excitation  melting lines for $q=0$ and $q=\Lambda_b$.  } 
\label{fig:alpha_evolution}
\end{figure}

\begin{figure}[ht!]
\centering
\begin{subfigure}{.5\textwidth}
\centering
\includegraphics[width=1.\linewidth]
{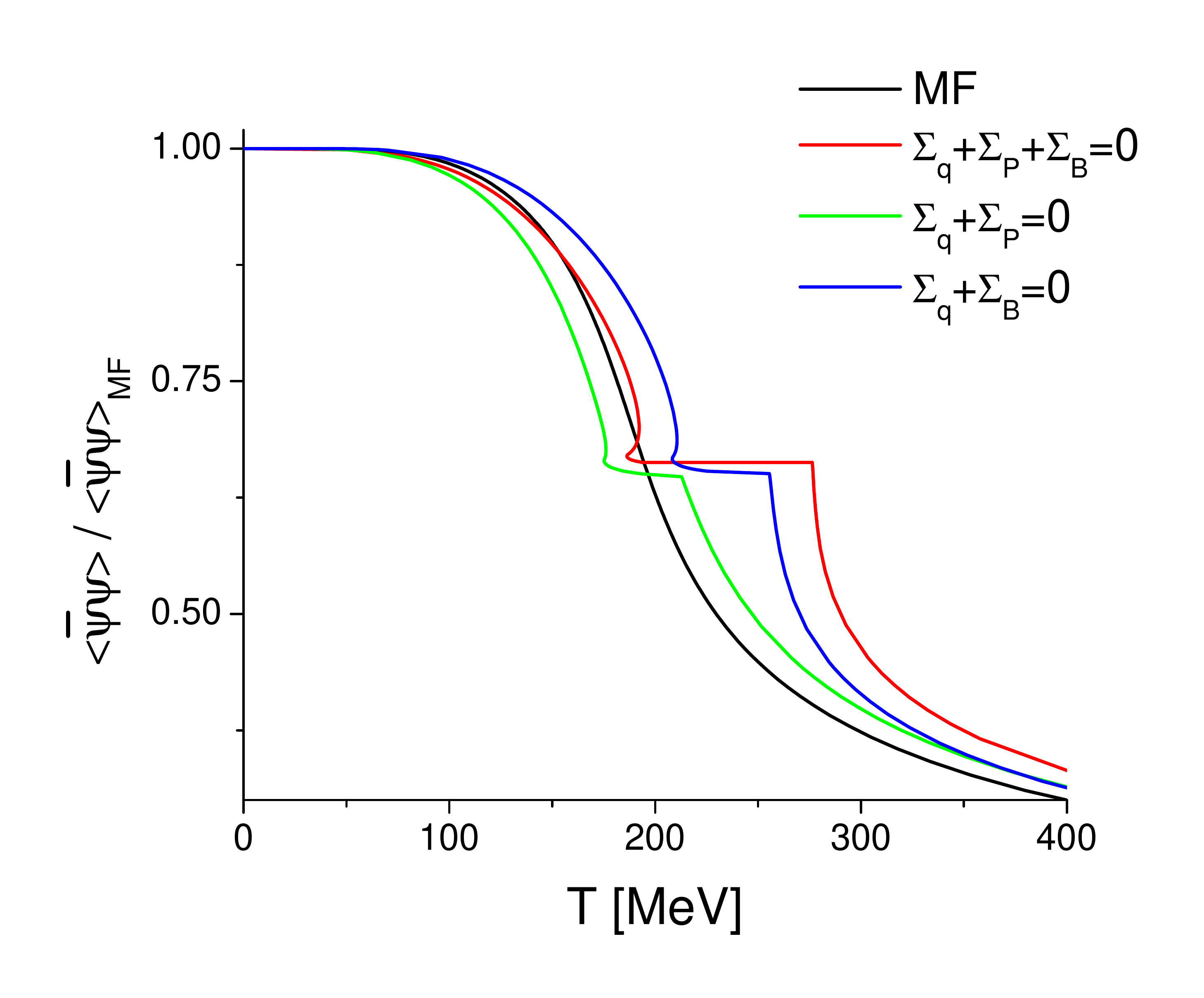}
\caption{\textit{Plasmon} limit, $\alpha=0.4$ }
\label{fig:separation_alpha_04_plasmon}
\end{subfigure}%
\begin{subfigure}{.5\textwidth}
\centering
\includegraphics[width=1.\linewidth]
{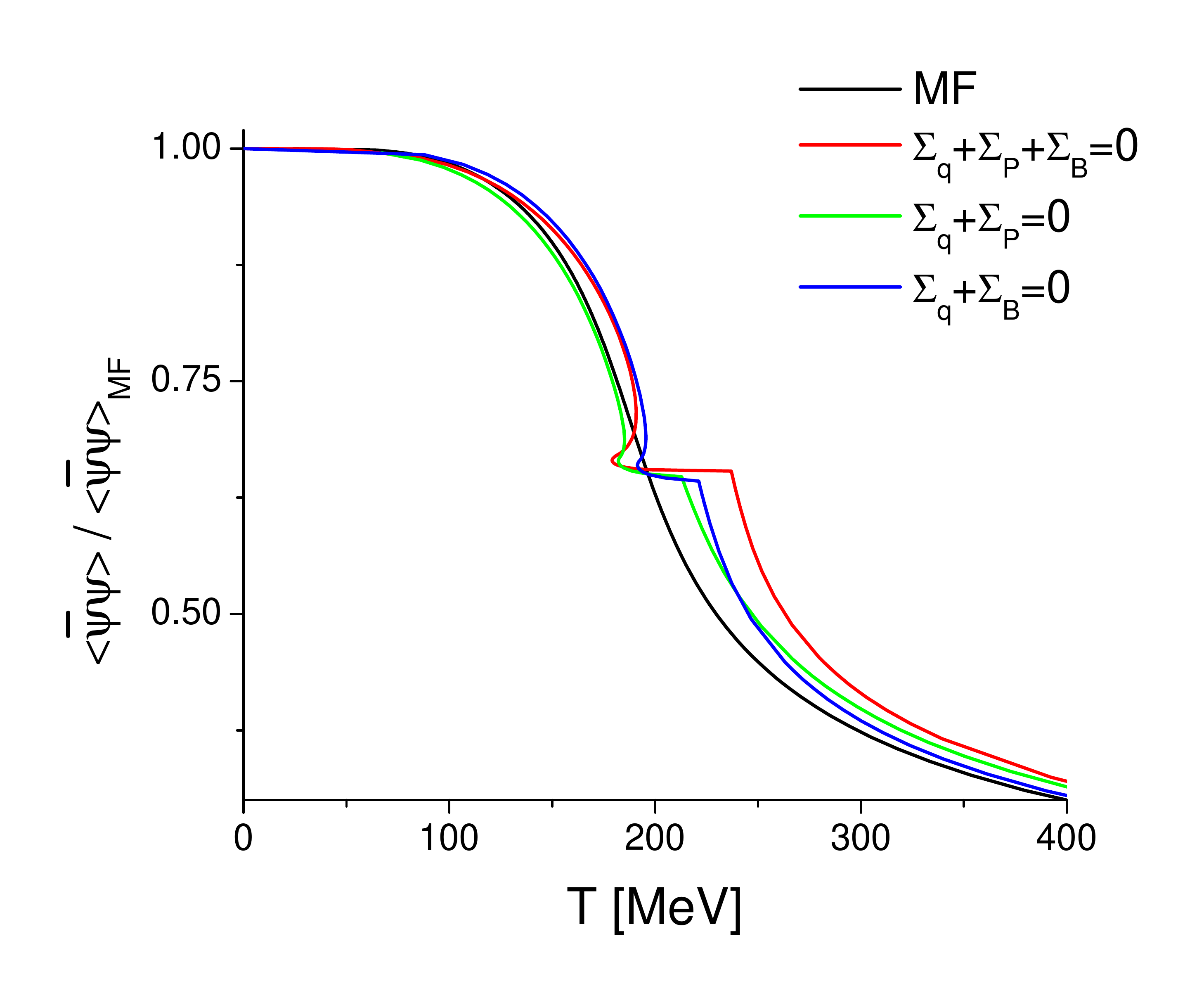}
\caption{\textit{Static} limit, $\alpha=0.4$ }
\label{fig:separation_alpha_04_static}
\end{subfigure}
\caption{\small Solution of the gap equation at finite temperature with $\alpha=\Lambda_b/\Lambda_f=0.4$. The left panel is the result in the \textit{plasmon} limit while the right panel is the \textit{static} limit. The green line is the result of solving the gap equation with the collective contributions, the blue line with the noncollective contributions and the red line is the complete calculation. The black line is the MF result using the parameters of Table \ref{MF_parameters}. }
\label{fig:gap_T_separating_contrubtions}
\end{figure}

When calculating the collective modes contributions to the gap equation at finite temperature, for a given pair of values $\qty(T,S)$, one is integrating over the meson momentum, from $0$ to $\Lambda_b$. However, as temperature increases, the value of $S$ decreases and chiral symmetry tends to get restored. As a consequence, the poles that originate the collective contributions and the branch cuts, move in the complex plane. Indeed, at a specific value of $\qty(T_0,S_0)$ the pole with momentum $q=0$, enters the branch cut (see Fig. \ref{fig:contour3}) and the mode with that dispersion relation no longer contributes as a collective excitation. As temperature continues to increase, more and more momentum modes generate pole contributions that overlap with the branch cuts and are not included as collective excitations. So, collective excitations are considered until the highest  boson momentum mode, with momentum $q=\Lambda_b$, enters the branch cut.

In Fig.  \ref{fig:alpha_evolution}, we present the results of solving the one meson loop gap equation, at finite temperature, increasing the boson cutoff. In all the panels we present the result of the MF model, using the parameters of Table \ref{MF_parameters}, for reference. We also present the so-called pion melting lines for pion collective modes with momentum $q=0$ and $q=\Lambda_b$ (dashed and full red lines of Fig.  \ref{fig:alpha_evolution}). For a given Hartree mass, these lines provide the respective melting temperature of the pion collective mode i.e., the temperature at which the poles with momentum modes $q=0$ and $q=\Lambda_b$, enter the branch cut. For $q=0$, this is known as the pion Mott temperature. The $q=0$ melting line, contrary to the $q=\Lambda_b$ one, depends only on the fermionic parameters i.e., it does not depend on the boson cutoff. This means that these lines are almost the same in all scenarios presented in Fig.  \ref{fig:alpha_evolution}. Upon solving these complete gap equation, once the quark condensate reaches this temperature, a smaller number of momentum modes will contribute to the collective modes.

In both the \textit{plasmon} and \textit{static} limits, the quark condensate at finite temperature, has a different behavior with meson loop corrections, when compared to MF, see Fig.  \ref{fig:gap_T_separating_contrubtions}. There is a bending behavior not seen at MF level: because of the inclusion of collective and non collective modes in the system and the crossing of cuts in the complex plane the quark condensate is not an analytical function of temperature. 

Figure \ref{fig:gap_T_separating_contrubtions} also shows that this behavior is present when solving the gap equation with both collective and non collective excitations (red line) or when considering these contributions separately (green and blue lines). Such observation leads us to conclude that this behavior is a consequence of considering beyond MF corrections within this formalism, independently if they are collective or non collective excitations.

Due to the presence of this bend, the critical temperature of the crossover transition cannot be defined as the zero of the second derivative of the quark condensate with respect to temperature, as usual. Still, one can clearly distinguish two phases, one with a large quark condensate and the other with a small quark condensate. These phases are also separated by the Mott temperature line of the $q=0$ pion collective mode (see Fig.  \ref{fig:alpha_evolution}). Hence, in this calculation, it would be natural to associate this temperature with the partial restoration of chiral symmetry.

A nonstandard quark condensate as a function of temperature was also obtained in \cite{Radzhabov:2010dd} for a nonlocal version of the Polyakov$-$Nambu$-$Jona-Lasinio model beyond mean field. In that work, the authors found a ``wiggle'' and attributed such a behavior to the beyond MF correction to the quark self-energy.

In conclusion, we expect to show that the inclusion of quantum fluctuations in the NJL model needs to be done with care, especially if one is trying to reproduce lattice QCD results. If one wants to have a consistent model beyond the mean field, including collective and noncollective excitations, one should also include in the gap equation contributions coming from these modes, as performed in this work. We found, however, that such a calculation leads to a strange behavior near the critical temperature of the model. This might indicate that the one-meson-loop NJL model, should be used with care as an effective model of QCD: taking our results into account, we conclude that to use the NJL with one-meson-loop corrections, the overall contribution from meson loops is very small i.e., $\alpha$, should be of the order of $\alpha=0.1-0.2$, to get a chiral condensate which is bounded by the error bars coming from 2-flavor lattice QCD calculations \cite{Karsch:2001cy}.

\section{Conclusions}
\label{conclusions}

In the present work, we have studied the effect of the inclusion of collective and noncollective modes in the quark condensate of the NJL model using a symmetry conserving approximation. This approximation is based on the effective action formalism and guarantees that the pion is the Goldstone mode in the chiral limit. 

Adding quantum fluctuations, in asymmetric conserving way, by considering the influence of collective and noncollective modes in the NJL model is not a simple task \cite{ZHUANG1994525}. The composite nature of the meson modes leads to a dynamical scenario where, depending on the temperature and Hartree mass, collective modes may, or may not exist. From the practical point of view, even evaluating some integrations analytically, one ends up effectively solving four dimensional integrals, numerically.

In the vacuum, using a mean field parametrization and adding the meson sector by increasing the boson cutoff, it was found a decreasing value for the quark condensate. This result is expected: the inclusion of boson degrees of freedom is known to drag the system into a state of restored chiral symmetry. 
It was also found that this decrease is limited by the existence of the collective modes. Decreasing the value of the condensate too much leads to the absence of pole contributions to the vacuum gap equation, which are essential to balance the gap equation, providing the existence of a solution, beyond the MF approximation.

This calculation shows that adding meson-loop correction terms to the NJL model, in a consistent way, is a very delicate process. There is a backreaction in the quark condensate and restoration of chiral symmetry, due to the existence of composite collective and noncollective modes. As temperature increases and chiral symmetry gets restored, the collective modes melt and its contribution to the gap equation vanishes.

As future work, testing the robustness of the results with different regularization procedures for the quark and meson loops, like the Pauli-Villars scheme, could be insightful. The calculation can be extended to finite density by including a finite chemical potential. With such an extension one could study in-medium behavior of the collective and noncollective modes and their influence on the restoration of chiral symmetry at finite density. This would also allow us to obtain the phase diagram of the NJL model at one-meson-loop level and check  the existence of a critical end-point and its robustness against increasing $\alpha=\Lambda_b/\Lambda_f$. Another interesting extension would be to include the Polyakov loop and study the influence of the collective and noncollective modes on the breaking of $Z\qty(N_c)$ symmetry and statistical deconfinement. 
Finally, the developed formalism can also be applied to the calculation of transport coefficients at finite temperature.

\section*{Acknowledgments}

The authors would like to thank Constança Providência, Hubert Hansen, Guy Chanfray and João Moreira for useful comments. This work was supported by national funds from FCT (Fundação para a Ciência e a Tecnologia, I.P, Portugal) under the IDPASC Ph.D. program (International Doctorate Network in Particle Physics, Astrophysics and Cosmology), with the Grant No. PD/BD/128234/2016 (R. C. P.), and under the Projects No. UID/FIS/04564/2016, No. UID/04564/2020, and No. POCI-01-0145-FEDER-029912 with financial support from POCI “Programa Operacional Competitividade e Internacionalização (COMPETE 2020)”, in its FEDER component.

\section{Appendix}
\label{appendix}

\subsection{Sokhotski-Plemelj formula}
\label{Plemelj_formula}

The Sokhotski-Plemelj formula is given by,
\begin{align}
\frac{1}{x-x_0 \pm i \epsilon} = \CPV \frac{1}{x-x_0} \mp i \pi \delta \qty( x-x_0 )  ,
\label{plemelj_formula}
\end{align}
where, $\epsilon \geq 0$ is an infinitesimal constant, $\CPV$ stands for the Cauchy principal value and $\delta$ is the Dirac delta function.

\subsection{ ${f_0(S)}$ loop function at finite temperature } %\bm
\label{appendix_f0_loop_function}

The thermal loop function $f_0 \qty(S)$ is defined as:
\begin{align}
f_0 \qty(S) & = 
\int_k
\frac{1}{ k^2 + S^2 } .
\end{align}
We can separate the time dependence by defining,
\begin{align}
E_{\vec{k}}^2 & = \vec{k}^2 + S^2 .
\label{def:Ek}
\end{align}
Writing the $k_0$ integrations as a sum over the fermionic Matsubara frequencies, $\omega_n = \nicefrac{\qty(2n+1)\pi}{\beta}, \; n \in \mathbb{Z}$ and using the contour integral technique to evaluate the sum, one gets:
\begin{align}
f_0 \qty(S) & =  
\int_{\vec{k}}
\frac{ 1 }{ 2 E_{\vec{k}} }
\qty[ 1 - 2 n_\fermi (E_{\vec{k}}) ] .
\end{align}

\subsection{ $f_1 \qty(S,q)$ loop function at finite temperature }
\label{appendix_f1_loop_function}

The thermal loop function $f_1 \qty(S,q)$ is defined as:
\begin{align*}
f_1 \qty(S,q) 
& =  \int_k
\frac{1}{ ( \qty(k-q)^2 + S^2 ) \qty( k^2 + S^2 )  } .
\numberthis
\end{align*}
We can separate the time dependence by using Eq.  (\ref{def:Ek}) and defining,
\begin{align}
E_{\vec{k}-\vec{q}}^2 & = \qty( \vec{k} - \vec{q} )^2 + S^2 .
\end{align}
To perform the integration over $k_0$, we write the integral as a sum over the allowed Matsubara frequencies, $\omega_n = \nicefrac{\qty(2n+1)\pi}{\beta}, \; n \in \mathbb{Z}$, for fermionic fields. The sum is then evaluated using the usual contour technique \cite{kapusta2006finite,Bellac:2011kqa}. This process will generate terms proportional to,
\begin{align}
n_\fermi \qty( iq_0 + E_{\vec{k}-\vec{q}} ) = \frac{ 1 }{ e^{ \beta iq_0 } e^{ \beta E_{\vec{k}-\vec{q}} } + 1 } ,
\end{align}
the Fermi distribution function with an external momentum $q_0$. This momentum corresponds to the the Matsubara frequency of an external particle. In this case, the external particles are bosons ($\sigma$ and $\vec{\pi}$ modes). Hence, $q_0 = \nicefrac{2n \pi}{\beta}$, $n \in \mathbb{Z}$. Making use of Euler's identity one writes,
\begin{align*}
n_\fermi \qty( iq_0 + E_{\vec{k}-\vec{q}} ) = 
\frac{ 1 }{ e^{ \beta iq_0 } e^{ \beta E_{\vec{k}-\vec{q}} } + 1 } = 
\frac{ 1 }{ e^{ \beta E_{\vec{k}-\vec{q}} } + 1 } =
n_\fermi \qty( E_{\vec{k}-\vec{q}} ) .
\end{align*}
After some calculations, one can finally arrive at,
\begin{align}
f_1\qty(S,\vec{q},q_0)  = 
\int_{\vec{k}}
\frac{1}{ 4 E_{\vec{k}} E_{\vec{k}-\vec{q}} } \Bigg\{
& 
\frac{ G_\P }{ iq_0 + E_\P }   - 
\frac{ G_\P }{ iq_0 - E_\P } 
+ 
\frac{ G_\M }{ iq_0 + E_\M }  -
\frac{ G_\M }{ iq_0 - E_\M } 
\Bigg\} ,
\label{eq:f1_k0_integrated}
\end{align}
where:
\begin{align}
E_\P & = E_{\vec{k}} + E_{\vec{k}-\vec{q}} ,
\\
E_\M  & = E_{\vec{k}} - E_{\vec{k}-\vec{q}} ,
\\
G_\P & = 1 - n_\fermi (E_{\vec{k}} ) - n_\fermi ( E_{\vec{k}-\vec{q}} ) ,
\\
G_\M & = n_\fermi (E_{\vec{k}} ) - n_\fermi ( E_{\vec{k}-\vec{q}} ) .
\end{align}

This function is nonanalytical at the origin, leading to two distinct results in the $q \to 0$ limit: the \textit{plasmon} and \textit{static} limits. For the \textit{plasmon} limit one can get:
\begin{align*}
\lim_{q_{0} \to 0} \lim_{\vec{q} \to 0} f_1 \qty(S,\vec{q},q_0) 
& =
\int \frac{ \dd[3]{k} }{(2\pi)^3} 
\frac{ 1 - 2 n_\fermi \qty( E_{\vec{k}} ) }{ 4 E_{\vec{k}}^3 }  .
\numberthis
\end{align*}
The \textit{static} limit can be calculate to yield, 
\begin{align*}
\lim_{\vec{q} \to 0} \lim_{q_{0} \to 0} f_1 \qty(S,\vec{q},q_0) 
& =
\int \frac{ \dd[3]{k} }{(2\pi)^3} 
\frac{ 1 }{ 4 E_{\vec{k}}^3 } 
\qty{
1 - 2n_\fermi (E_{\vec{k}})
+ 
\frac{ 2 E_{\vec{k}} }{ T }
n_\fermi (E_{\vec{k}})
\qty[  
n_\fermi (E_{\vec{k}}) - 1
]  
} .
\numberthis
\end{align*}
Both expressions agree in the zero temperature limit, as expected. For more details see \cite{das1997finite}.

In the calculations we are interested in the function $f_1\qty(S,\vec{q},-i\omega)$ with $q_0$ a pure imaginary number. Consider a Wick rotation $q_0=-i\omega$, for real $\omega$ and define $F\qty(S,\vec{q},\omega)$ as:
\begin{align*}
F\qty(S,\vec{q},\omega) & = 
f_1\qty(S,\vec{q},-i\omega) 
\\
& =
\int_{\vec{k}}
\frac{1}{ 4 E_{\vec{k}} E_{\vec{k}-\vec{q}} } 
\qty{
\frac{ G_\P }{ \omega + E_\P }   - 
\frac{ G_\P }{ \omega - E_\P } 
+ 
\frac{ G_\M  }{ \omega + E_\M }  -
\frac{ G_\M  }{ \omega - E_\M } 
} .
\numberthis
\label{eq:F(S,q,w)_definition}
\end{align*}
The real and imaginary parts of $F \qty(S,\vec{q}, \omega)$ can be defined, near the real axis, with an analytical continuation. Following \cite{YAMAZAKI201319,YAMAZAKI2014237}, we write:
\begin{align}
F \qty(S,\vec{q}, \omega) \to F \qty(S,\vec{q}, \omega \pm i\epsilon) & =
\Re \qty[ F \qty(S,\vec{q}, \omega) ] \pm i \Im \qty[ F \qty(S,\vec{q}, \omega) ] .
\label{eq:F_real_im_definiton}
\end{align}
The function $F \qty(S,\vec{q}, \omega)$ in Eq.  (\ref{eq:F(S,q,w)_definition}), is an even function with respect to $\omega$ i.e., $F \qty(S,\vec{q}, \omega \pm i\epsilon ) = F \qty(S,\vec{q}, -\qty(\omega \pm i\epsilon)) $. By defining the real and imaginary parts as in Eq.  (\ref{eq:F_real_im_definiton}), this property implies that, near the real axis, the real part will be an even function of $\omega$ while, the imaginary part will be an odd function \cite{Peskin:1995ev}. Indeed one can write:
\begin{align}
\Re \qty[ F \qty(S,\vec{q}, \omega) ] & = \Re \qty[ F \qty(S,\vec{q}, -\omega) ] ,
\label{def:ReF_even}
\\
\Im \qty[ F \qty(S,\vec{q}, \omega) ] & = - \Im \qty[ F \qty(S,\vec{q}, -\omega) ] .
\label{def:ImF_odd}
\end{align}
Each contribution defined in Eq. (\ref{eq:F_real_im_definiton}) can be explicitly calculated by applying the Sokhotski-Plemelj formula for distributions defined in Eq.  (\ref{plemelj_formula}).

%%%%%%%%%%%%%%%%%%%%%%%%%%%%%%%%%%%%%%%%%%%%%%%%%%%%%%%%%%%%%%%%%%%%%%%%%%%%%%%%%%%%%%%%%%%%%%%%
%%%%%%%%%%%%%%%%%%%%%%%%%%%%%%%%%%%%%%%%%%%%%%%%%%%%%%%%%%%%%%%%%%%%%%%%%%%%%%%%%%%%%%%%%%%%%%%%

\subsection{ The $I_M \qty(S)$ contribution}
\label{appendix_IM}

Consider the term given in Eq.  (\ref{def:IM}), for a given meson channel $M=\{ \sigma, \pi \}$. We can write it as:
\begin{align*}
I_M \qty(S) & = 
\frac{1}{ 2 N_c N_f }
\int_{ \vec{q} }
\int\frac{ \dd{q_0} }{ 2\pi }
\qty[ f_1\qty(S,\vec{q},q_0) k_M^{-1} \qty(S,\vec{q}, q_0 ) + \tilde{m} ]^{-1} .
\numberthis
\end{align*}
Changing the integration over $q_0$ into a sum over Matsubara frequencies $\omega_n$, one gets,
\begin{align}
I_M \qty(S) & = 
\frac{1}{ 2 N_c N_f }
\int_{ \vec{q} }
\frac{1}{\beta}
\sumMatsubara
\qty[ f_1\qty(S,\vec{q},\omega_n) k_M^{-1} \qty(S,\vec{q}, \omega_n ) + \tilde{m} ]^{-1}  .
\end{align}
As already stated, $q$ corresponds to the momentum of a composite boson hence, $\omega_n=\frac{ 2 n \pi }{ \beta }$, the bosonic Matsubara frequencies. This sum can be converted into a contour integration, using contour $\mathcal{C}$ of Fig.  \ref{fig:contour2}. One gets,
\begin{align}
I_M \qty(S) 
& = 
\frac{1}{ 2 N_c N_f }
\int_{ \vec{q} }
\frac{1}{2}
\oint_{\mathcal{C}}
\frac{ \dd{w} }{ 2 \pi i }
\coth \qty( \frac{ \beta w }{ 2 } )
\qty[ f_1\qty(S,\vec{q},-i w) k_M^{-1} \qty(S,\vec{q},-i w) + \tilde{m} ]^{-1} 
.
\end{align}
Applying the formalism discussed earlier through Eq.  (\ref{eq:general_contour_integration}), the contour integral can be converted into an integration around the real axis, in which only the imaginary part of the integrand will contribute to the final result. The integral can then be divided in the collective and noncollective contributions as indicated in Eq.  (\ref{def:IM_pole_cut_decomposition}).

The first term, $\mathcal{P}_M \qty(S)$, can be calculated by considering that, near the real axis, the loop function $f_1\qty(S,\vec{q},-i w)$ is purely real and $k_M^{-1} \qty(S,\vec{q},-i w)$ as an imaginary part. One can write,
\begin{align*}
\mathcal{P}_M \qty(S) 
& = 
\frac{1}{ 4 \pi N_c N_f }
\int_{ \vec{q} }
\int_{-\infty}^{+\infty} \dd{\omega}
\frac{ \coth \qty( \nicefrac{ \beta \omega }{ 2 } ) }{ \Re \qty[ F \qty(S,\vec{q}, \omega) ] }
\Im \qty[ 
\qty(
K_M \qty(S,\vec{q},\omega + i\epsilon )^{-1} + 
\frac{ \tilde{m} }{ \Re \qty[ F \qty(S,\vec{q}, \omega) ] }
)^{-1} 
] .
\numberthis
\end{align*}
Using the Sokhotski-Plemelj formula and the properties of the Dirac delta function, the imaginary part of the integrand is,
\begin{align}
\Im \qty[ 
\qty(
K_M \qty(S,\vec{q},\omega + i\epsilon )^{-1} + 
\frac{ \tilde{m} }{ \Re \qty[ F \qty(S,\vec{q}, \omega) ] }
)^{-1} 
]
& =
\frac{ \pi }{ 2 \tilde{E}_M (S,\vec{q},\omega) }
\sumeta
\eta
\frac{ \delta \qty( \omega - \omega_\eta ) }{ \abs{ \partial_\omega \chi_\eta \qty(S,\vec{q},\omega ) }  }_{\omega_\eta}  ,
\label{eq:IM_integrand}
\end{align}
where the quantity $\chi_\eta \qty(S,\vec{q},\omega)$ and its $\omega$-derivative are given by:
\begin{align}
\chi_\eta \qty(S,\vec{q},\omega) & = \omega - \eta \tilde{E}_M (S,\vec{q},\omega) ,
\\
\partial_\omega \chi_\eta \qty(S,\vec{q},\omega) 
& = 
1 + 
\frac
{ \eta  \tilde{m} }
{ 2 \tilde{E}_M (S,\vec{q},\omega) } 
\frac
{ \partial_\omega \Re \qty[ F \qty(S,\vec{q}, \omega) ] }
{ \qty(\Re \qty[ F \qty(S,\vec{q}, \omega) ])^2 } ,
\end{align}
with $\omega_\eta$ a solution to Eq.  (\ref{def:pole_location_eq}). Plugging the imaginary part in the integral, and using the delta function to integrate over $\omega$ yields the final result:
\begin{align*}
\mathcal{P}_M \qty(S) 
& = 
\frac{1}{ 4 N_c N_f }
\int_{ \vec{q} }
\frac{ \coth \qty( \nicefrac{ \beta \omega_\P }{ 2 } ) }{ \Re \qty[ F \qty(S,\vec{q}, \omega_\P ) ] }
\frac{ \abs{ \partial_\omega \chi_\P \qty(S,\vec{q},\omega ) }_{\omega_\P}^{-1} }{ \tilde{E}_M (S,\vec{q}, \omega_\P ) }   .
\numberthis
\end{align*}

Considering that $k_M^{-1} \qty(S,\vec{q},-i w)$ is real while $f_1\qty(S,\vec{q},-i w)$ is complex will give the branch cut contribution. One can write,
\begin{align}
\mathcal{B}_M \qty(S) 
& = 
\frac{1}{ 4 \pi N_c N_f }
\int_{ \vec{q} }
\int_{-\infty}^{+\infty} \dd{\omega}
\frac{ \coth \qty( \nicefrac{ \beta \omega }{ 2 } ) }{ -\omega^2 + { {E} }^2_M \qty(S,\vec{q}) }
\Im \qty[ 
\qty(
F \qty(S,\vec{q}, \omega)  + \tilde{m} K_M \qty(S,\vec{q},\omega ) 
)^{-1} 
]
.
\end{align}
Using Eq.  (\ref{eq:def_ReG}), near the real axis, the quotient in the integrand can be written as,
\begin{align*}
\qty[
F \qty(S,\vec{q}, \omega + i \epsilon ) + \tilde{m} K_M \qty(S,\vec{q},\omega ) 
]^{-1} 
= \frac{ \Re \qty[ G \qty(S,\vec{q}, \omega ) ] - i \Im \qty[ F \qty(S,\vec{q}, \omega ) ] }{ \Re \qty[ G \qty(S,\vec{q}, \omega ) ]^2 + \Im \qty[ F \qty(S,\vec{q}, \omega ) ]^2  }  .
\end{align*}
One can drop the real part of this expression and write,
\begin{align}
\mathcal{B}_M \qty(S) 
& =
\frac{1}{ 2 \pi N_c N_f }
\int_{ \vec{q} }
\int_{ 0 }^{+\infty} \dd{\omega}
\frac{ \coth \qty( \nicefrac{ \beta \omega }{ 2 } ) }{ -\omega^2 + { {E} }^2_M \qty(S,\vec{q}) }
\frac{ - \Im \qty[ F \qty(S,\vec{q}, \omega ) ]  }{ \Re \qty[ G \qty(S,\vec{q}, \omega ) ]^2 + \Im \qty[ F \qty(S,\vec{q}, \omega ) ]^2 }
,
\end{align}
where we used the fact that the integrand is even in $\omega$.

%%%%%%%%%%%%%%%%%%%%%%%%%%%%%%%%%%%%%%%%%%%%%%%%%%%%%%%%%%%%%%%%%%%%%%%%%%%%%%%%%%%%%%%%%%%%%%%%
%%%%%%%%%%%%%%%%%%%%%%%%%%%%%%%%%%%%%%%%%%%%%%%%%%%%%%%%%%%%%%%%%%%%%%%%%%%%%%%%%%%%%%%%%%%%%%%%

\subsection{ The $I_{1M} \qty(S)$ contribution}
\label{appendix_I1M}

As already stated, only the branch cut contribution of the $I_{1\sigma} \qty(S)$ integral needs to be calculated. Consider,
\begin{align}
I_{1\sigma} \qty(S)
& =
\frac{1}{ 2 N_c N_f }
\int_{ \vec{q} }
\int\frac{ \dd{q_0} }{ 2\pi }
f_1\qty(S,\vec{q},q_0)
\qty[ f_1\qty(S,\vec{q},q_0) k_\sigma^{-1} \qty(S,\vec{q}, q_0 ) + \tilde{m} ]^{-1}   .
\end{align}
By changing the integral into a Matsubara sum and then to a contour integration using contour $\mathcal{C}$, one gets:
\begin{align*}
I_{1\sigma} \qty(S) 
& = 
\frac{1}{ 2 N_c N_f }
\int_{ \vec{q} }
\frac{1}{2}
\oint_{\mathcal{C}}
\frac{ \dd{w} }{ 2 \pi i }
\coth \qty( \frac{ \beta w }{ 2 } )
f_1\qty(S,\vec{q},-i w)
\qty[ f_1\qty(S,\vec{q},-i w) k_\sigma^{-1} \qty(S,\vec{q}, -i w ) + \tilde{m} ]^{-1}   .
\numberthis
\end{align*}
Following the usual recipe to calculate the noncollective mode contribution, $\mathcal{B}_{1\sigma} \qty(S)$ is given by,
\begin{align}
\mathcal{B}_{1\sigma} \qty(S)
& = 
\frac{1}{ 4 \pi N_c N_f }
\int_{ \vec{q} }
\int_{-\infty}^{+\infty} \dd{\omega}
\frac{ \coth \qty( \nicefrac{ \beta \omega }{ 2 } ) }{ -\omega^2 + { {E} }^2_\sigma \qty(S,\vec{q}) }
\Im \qty[ 
\frac{ F \qty(S,\vec{q}, \omega) }
{ F \qty(S,\vec{q}, \omega)  + \tilde{m} K_M \qty(S,\vec{q},\omega )  }
]  .
\end{align}
Near the real axis, one can write:
\begin{align}
\frac{ F \qty(S,\vec{q}, \omega + i\epsilon ) }
{ F \qty(S,\vec{q}, \omega + i\epsilon )  + \tilde{m} K_M \qty(S,\vec{q},\omega ) }
& = 
1
- \tilde{m} K_M \qty(S,\vec{q},\omega )
\frac{ \Re \qty[ G \qty(S,\vec{q}, \omega ) ] - i \Im \qty[ F \qty(S,\vec{q}, \omega ) ] }{ \Re \qty[ G \qty(S,\vec{q}, \omega ) ]^2 + \Im \qty[ F \qty(S,\vec{q}, \omega ) ]^2  }  .
\end{align}
Considering only the imaginary part of the above quotient, one gets:
\begin{align}
\mathcal{B}_{1\sigma} \qty(S)
& = 
\frac{ \tilde{m} }{ 2 \pi N_c N_f }
\int_{ \vec{q} }
\int_{ 0 }^{+\infty} \dd{\omega}
\frac{ \coth \qty( \nicefrac{ \beta \omega }{ 2 } ) }{ -\omega^2 + { {E} }^2_\sigma \qty(S,\vec{q}) }
\frac{ K_\sigma \qty(S,\vec{q},\omega ) \Im \qty[ F \qty(S,\vec{q}, \omega ) ] }{ \Re \qty[ G \qty(S,\vec{q}, \omega ) ]^2 + \Im \qty[ F \qty(S,\vec{q}, \omega ) ]^2  } .
\end{align}

%%%%%%%%%%%%%%%%%%%%%%%%%%%%%%%%%%%%%%%%%%%%%%%%%%%%%%%%%%%%%%%%%%%%%%%%%%%%%%%%%%%%%%%%%%%%%%%%
%%%%%%%%%%%%%%%%%%%%%%%%%%%%%%%%%%%%%%%%%%%%%%%%%%%%%%%%%%%%%%%%%%%%%%%%%%%%%%%%%%%%%%%%%%%%%%%%

\subsection{ The $I_{2M} \qty(S)$ contribution}
\label{appendix_I2M}

The final and more complicated contribution comes from integrals $I_{2\sigma} \qty(S)$ and $I_{2\pi} \qty(S)$. We can define the quantity, $I_{2M} \qty(S)$, which depends on the meson channel $M=\{ \sigma, \pi \}$ as:
\begin{align*}
I_{2M} \qty(S) & = - 2 \int\frac{ \dd[4]{q} }{\qty(2\pi)^4} \qty( q^2 + 4S^2 \delta_{ M\sigma } )f_2\qty(S,q) \tilde{ \Delta }_M (S,q) .
\end{align*}
To simplify the calculations, we use the identity presented in Eq.  (\ref{eq:f2_f1_relation}) and write,
\begin{align}
I_{2M} \qty(S) 
& =
\frac{1}{ 2 N_c N_f }
\int_{ \vec{q} }
\pdv{ \xi^2 }
\int\frac{ \dd{q_0} }{ 2\pi }
\ln
\qty{
f_1 \qty(\xi,\vec{q},q_0) k_M^{-1} \qty(S,\vec{q}, q_0 ) + \tilde{m}
}
_{\xi=S}  .
\end{align}
Here, the $\xi^2$ derivative commutes with the integration since the integral bounds are $\xi$-independent. Following the usual recipe, the $q_0$ integration can be transformed into a Matsubara sum. The sum is then converted into a contour integration, using contour $\mathcal{C}$. We can now write,
\begin{align}
I_{2M} \qty(S) 
& =
\frac{1}{ 2 N_c N_f }
\int_{ \vec{q} }
\pdv{ \xi^2 }
\frac{1}{2}
\oint_{\mathcal{C}}
\frac{ \dd{w} }{ 2 \pi i }
\coth \qty( \frac{ \beta w }{ 2 } )
\ln
\qty{
f_1 \qty(\xi,\vec{q},-i w) k_M^{-1} \qty(S,\vec{q}, -i w ) + \tilde{m}
}
_{\xi=S}  .
\end{align}
The separation in the pole and branch cut contributions is performed using Eq.  (\ref{def:I2M_P2M_plus_B2M}).

For the pole contribution $\mathcal{P}_{2M} \qty(S)$, one can write:
\begin{align}
\mathcal{P}_{2M} \qty(S)
& =
\frac{1}{ 4 \pi N_c N_f }
\int_{ \vec{q} }
\pdv{ \xi^2 }
\int_{-\infty}^{+\infty} \dd{\omega}
\coth \qty( \frac{ \beta \omega }{ 2 } )
\Im \qty[
\ln
\qty{
\Re \qty[ F \qty(\xi,\vec{q}, \omega) ] K_M^{-1} \qty(S,\vec{q}, \omega ) + \tilde{m}
}
]
_{\xi=S} .
\end{align}
The logarithm in the integrand can be written as,
\begin{align}
\ln
\qty{
\Re \qty[ F \qty(\xi,\vec{q}, \omega) ] K_M^{-1} \qty(S,\vec{q}, \omega ) + \tilde{m}
}
& =
\ln
\Re \qty[ F \qty(\xi,\vec{q}, \omega) ]
+
\ln
\qty{
K_M^{-1} \qty(S,\vec{q}, \omega ) + \frac{ \tilde{m} }{ \Re \qty[ F \qty(\xi,\vec{q}, \omega) ] } 
} .
\end{align}
The first term is real and can be dropped. Hence,
\begin{align*}
\mathcal{P}_{2M} \qty(S)
& =
\frac{1}{ 4 \pi N_c N_f }
\int_{ \vec{q} }
\pdv{ \xi^2 }
\int_{-\infty}^{+\infty} \dd{\omega}
\coth \qty( \frac{ \beta \omega }{ 2 } )
\Im \qty[
\ln
\qty{
K_M^{-1} \qty(S,\vec{q}, \omega ) + \frac{ \tilde{m} }{ \Re \qty[ F \qty(\xi,\vec{q}, \omega) ] } 
}
]
_{\xi=S} .
\end{align*}
Calculating the derivative yields:
\begin{align*}
\mathcal{P}_{2M} \qty(S)
& =
\frac{1}{ 4 \pi N_c N_f }
\int_{ \vec{q} }
\int_{-\infty}^{+\infty} \dd{\omega}
\coth \qty( \frac{ \beta \omega }{ 2 } )
\Im \qty[
\qty{
K_M^{-1} \qty(S,\vec{q}, \omega ) + \frac{ \tilde{m} }{ \Re \qty[ F \qty(S,\vec{q}, \omega) ] } 
}^{-1}
]
\pdv{ \xi^2 }
\frac{ \tilde{m} }{ \Re \qty[ F \qty(\xi,\vec{q}, \omega) ] }
_{\xi=S}  .
\end{align*}
Using Eq. (\ref{eq:IM_integrand}) and defining $R_F \qty(S,\vec{q}, \omega) = 16 \pi^2 q \Re \qty[ F \qty(S,\vec{q}, \omega)]$, the final result is given by:
\begin{align}
\mathcal{P}_{2M} \qty(S)
& =
-
\frac{ 4 \pi^2 \tilde{m} }{ N_c N_f }
\int_{ \vec{q} }
q 
\frac{ \coth \qty( \nicefrac{ \beta \omega_\P }{ 2 } ) }{ \tilde{E}_M (S,\vec{q},\omega_\P) }
\frac{ \partial_{S^2} R_F \qty(S,\vec{q}, \omega_\P) }{ R_F \qty(S,\vec{q}, \omega_\P)^2 }
\abs{ \partial_\omega \chi_\P \qty(S,\vec{q},\omega ) }_{\omega_\P}^{-1}  .
\end{align}

For the branch cut contribution $\mathcal{B}_{2M} \qty(S)$, consider,
\begin{align*}
\mathcal{B}_{2M} \qty(S)
& =
\frac{1}{ 4 \pi N_c N_f }
\int_{ \vec{q} }
\pdv{ \xi^2 }
\int_{-\infty}^{+\infty} \dd{\omega}
\coth \qty( \frac{ \beta \omega }{ 2 } )
\Im \qty[
\ln
\qty{
F \qty(\xi,\vec{q}, \omega) K_M^{-1} \qty(S,\vec{q}, \omega ) + \tilde{m}
}
]
_{\xi=S} \,.
\numberthis
\end{align*}
The logarithm can be written as:
\begin{align}
\ln
\bigg\{
F \qty(\xi,\vec{q}, \omega) K_M^{-1} \qty(S,\vec{q}, \omega ) + \tilde{m}
\bigg\}
& =
-\ln
K_M \qty(S,\vec{q}, \omega )
+
\big\{  
F \qty(\xi,\vec{q}, \omega) + \tilde{m} K_M \qty(S,\vec{q},\omega)
\big\} .
\end{align}
The first term can be dropped since it a pure real number and $\mathcal{B}_{2M} \qty(S)$ can be written as:
\begin{align*}
\mathcal{B}_{2M} \qty(S)
& =
\frac{1}{ 4 \pi N_c N_f }
\int_{ \vec{q} }
\pdv{ \xi^2 }
\int_{-\infty}^{+\infty} \dd{\omega}
\coth \qty( \frac{ \beta \omega }{ 2 } )
\Im \qty[
\ln
\bigg\{  
F \qty(\xi,\vec{q}, \omega) + \tilde{m} K_M \qty(S,\vec{q},\omega)
\bigg\}
]
_{\xi=S} \,.
\numberthis
\end{align*}
To calculate this term, the definition of $\Re \qty[ G \qty(\xi,\vec{q}, \omega ) ]$ is slightly different from the one in Eq.  (\ref{eq:def_ReG}). The term coming from $\tilde{m} K_M \qty(S,\vec{q},\omega)$ does not depend on $\xi$,
\begin{align}
\Re \qty[ G \qty(\xi,\vec{q}, \omega ) ] 
& =
\Re \qty[F \qty(\xi,\vec{q}, \omega)] 
+ M \qty(S,\vec{q},\omega ) .
\end{align}
The argument of the logarithm, near the real axis, can be written as:
\begin{align*}
F \qty(\xi,\vec{q}, \omega + i \epsilon) + \tilde{m} K_M \qty(S,\vec{q},\omega)
& =
\Re \qty[ G \qty(\xi,\vec{q}, \omega ) ] + i \Im \qty[F \qty(\xi,\vec{q}, \omega)] .
\end{align*}
The real part of the function $F \qty(S,\vec{q},\omega)$ is even and its imaginary part is odd, with respect to $\omega$. Using these properties, the integration is broken at $\omega = 0$ and a variable change in the integration for negative $\omega$ as $\omega=-\omega$, provides:
\begin{align*}
\mathcal{B}_{2M} \qty(S)
& =
\frac{1}{ 4 \pi N_c N_f }
\int_{ \vec{q} }
\pdv{ \xi^2 }
\int_{0}^{+\infty} \dd{\omega}
\coth \qty( \frac{ \beta \omega }{ 2 } )
\Im \qty[
\ln
\qty{  
\frac
{ \Re \qty[ G \qty(\xi,\vec{q}, \omega ) ] + i \Im \qty[F \qty(\xi,\vec{q}, \omega)] }
{ \Re \qty[ G \qty(\xi,\vec{q}, \omega ) ] - i \Im \qty[F \qty(\xi,\vec{q}, \omega)]}
}
]
_{\xi=S}  .
\end{align*}
The complex numbers in the logarithm argument can be written in the polar representation by defining their absolute value $L \qty(\xi,\vec{q}, \omega)$ and argument $A \qty(\xi,\vec{q}, \omega)$ as,
\begin{align}
L \qty(\xi,\vec{q}, \omega) & = 
\sqrt{ 
\Re \qty[ G \qty(\xi,\vec{q}, \omega ) ]^2 + 
\Im \qty[F \qty(\xi,\vec{q}, \omega)]^2 
}  , 
\label{eq:F_modulo}
\\
A \qty(\xi,\vec{q}, \omega) & = 
\frac
{ \Im \qty[F \qty(\xi,\vec{q}, \omega)] }
{ \Re \qty[ G \qty(\xi,\vec{q}, \omega ) ] }  ,
\label{eq:F_arg}
\end{align}
which allows to write,
\begin{align*}
\Re \qty[ G \qty(\xi,\vec{q}, \omega ) ] \pm i \Im \qty[F \qty(\xi,\vec{q}, \omega)] 
& =
L \qty(\xi,\vec{q}, \omega) 
\exp \Big[ \pm i \arctg A \qty(\xi,\vec{q}, \omega) \Big] .
\end{align*}
Using the polar representation and commuting the $\xi^2$ derivative with the $\omega$ integral, it gives:
\begin{align}
\mathcal{B}_{2M} \qty(S)
& =
\frac{1}{ 2 \pi N_c N_f }
\int_{ \vec{q} }
\int_{0}^{+\infty} \dd{\omega}
\frac{ \coth \qty( \nicefrac{ \beta \omega }{ 2 } ) }{1 + A \qty(S,\vec{q}, \omega)^2} 
\partial_{\xi^2} A \qty(\xi,\vec{q}, \omega)_{\xi=S}  \,.
\end{align}

\end{document}